\documentclass[twocolumn,floatfix,prb,aps,showpacs,bibliography]{revtex4-1}
\usepackage{graphicx,color,url,nicefrac,color,multirow,amsmath,amssymb,amsfonts,graphicx,tabularx}
\usepackage[titletoc,title]{appendix}
\usepackage[titletoc,title]{appendix}
\usepackage[unicode=true,colorlinks=true]{hyperref}
\hypersetup{linkcolor=blue,citecolor=blue,urlcolor=blue}
\makeatletter
\newcommand*{\rom}[1]{\expandafter\@slowromancap\romannumeral #1@}
\makeatother

\newcommand{\ie}{\emph{i.e.}\,}

\begin{document}

\newcommand{\CalF}{{\mathcal{F}}}
\newcommand{\NSymPol}{N}
\newcommand{\NSDTotal}{N}
\newcommand{\NSDGroup}{\mathcal{N}}
\newcommand{\Nsphere}{{{N}}}
\newcommand{\Nb}{{{N_b}}}
\newcommand{\LAngMom}{{\mathcal{L}}}
\newcommand{\LLockZero}{{L}}

\renewcommand*\arraystretch{1.5} 

\title{Search for exact local Hamiltonians for general fractional quantum Hall states}
\author{G J Sreejith$^{1}$, M Fremling$^{2,3}$, Gun Sang Jeon$^{4,5}$ and J K Jain$^{5}$}
\affiliation{$^{1}$ Indian Institute of Science Education and Research, Pune 411008 India}
\affiliation{$^{2}$Department of Theoretical Physics, Maynooth University, Maynooth, Co.Kildare, W23 HW31, Ireland}
\affiliation{$^{3}$Institute for Theoretical Physics, Center for Extreme Matter and Emergent Phenomena,
Utrecht University, Princetonplein 5, 3584 CC Utrecht, the Netherlands}
\affiliation{$^{4}$Department of Physics, Ewha Womans University, Seoul 03760, Korea}
\affiliation{$^{5}$Department of Physics, 104 Davey Lab, Pennsylvania State University, University Park, Pennsylvania 16802, USA}

\date{\today}

\begin{abstract}
We report on our systematic attempts at finding local interactions for which the lowest-Landau-level projected composite-fermion wave functions are the unique zero energy ground states. For this purpose,
we study in detail the simplest non-trivial system beyond the Laughlin states,
namely bosons at filling factor $\nu=\nicefrac{2}{3}$ and identify local constraints among clusters of particles in the ground state. By explicit calculation,
we show that no Hamiltonian up to (and including) four particle interactions produces this state as the exact ground state,
and speculate that this remains true even when interaction terms involving greater number of particles are included. Surprisingly, we can  identify an interaction,
which imposes an energetic penalty for a specific entangled configuration of four particles with relative angular momentum of $6\hbar$,
that produces a unique zero energy solution (as we have confirmed for up to 12 particles). This state, referred to 
as the $\lambda$-state, is not identical to the projected composite-fermion state, but the following facts suggest that the two might be topologically equivalent: the two sates have a high overlap; they have the same root partition; the quantum numbers for their neutral excitations are identical; and the quantum numbers for the quasiparticle excitations also match. {On the quasihole side, we find that even though the quantum numbers of the lowest energy states agree with the prediction from the composite-fermion theory, these states are not separated from the others by a clearly identifiable gap. This prevents us from making a conclusive claim regarding the topological equivalence of the $\lambda$ state and the composite-fermion state.}
Our study illustrates how new candidate states can be identified from constraining selected many particle configurations 
and it would be interesting to pursue their topological classification.
\end{abstract}

\maketitle

\section{Introduction}
Exactly solvable models that capture nontrivial physics of real systems occupy a special place in physics.
The quest for such models for strongly correlated fractional quantum Hall states began with Haldane's construction\cite{Haldane83}
of a truncated pseudopotential interaction that obtains the Laughlin wave function\cite{Laughlin83} at filling factor $\nu=\nicefrac{1}{(m+1)}$ as the exact zero energy state.
(Here $m$ is an even integer for fermions and an odd integer for bosons.) The model interaction (see also Ref.~\onlinecite{Trugman85}) was reverse-engineered from the observation that no pair of electrons has a relative angular momentum $m$ or less in the Laughlin wave function,
and imposing an energetic penalty for these pairs produces the Laughlin's state uniquely.
While model Hamiltonians with multi-particle interaction have also been constructed for certain simple,
\ie naturally holomorphic wave functions,
no interactions have been identified that yield the lowest Landau level (LLL) projected Jain wave functions\cite{Jain89,Jain90,Jain07} at fractions $\nu=n/(mn\pm 1)$ as the unique exact solutions.
These reduce to the Laughlin wave function for $n=1$,
but the pathway of constructing the wave functions for $n> 1$ passes through higher Landau levels (LLs).
In their simplest physically-transparent ``unprojected" form,
these wave functions are non-holomorphic, with a small finite amplitude spilling outside the LLL,
and projection into the LLL yields a rather complicated wave function tangled by the presence of strategically placed derivatives.
While these LLL wave functions can be evaluated on the computer and turn out to be extremely close to the exact Coulomb ground states,
the final form of the wave function is too complicated to afford a hint into the construction of exact parent Hamiltonians.

It is worth stressing that the LLL constraint, while undoubtedly useful for testing theories against computer experiments,
is not a necessary condition for the occurrence of fractional quantum Hall effect (FQHE). In real experiments,
the fractional quantum Hall states always involve some Landau level (LL) mixing,
which provides no correction to the quantized Hall resistance (unless it is large enough to close the gap).
The unprojected Jain wave functions are sufficiently simple to allow, in some instances,
construction of exact parent Hamiltonians in which some of the lowest LLs are made degenerate while the rest are sent to infinite energies\cite{Jain90,Jain89b,Bandyopadhyay18,ChenLi2017}.  In such cases,
an adiabatic path connecting the model Hamiltonian to the LLL Coulomb Hamiltonian may be identified along which the gap does not close,
thus establishing the adiabatic continuity between the unprojected and the LLL projected wave functions\cite{Rezayi91}. One may take the view, however,
that the LLL limit, while not necessary, is an interesting one,
and local interactions that directly produce the LLL projected wave functions as exact solutions may offer important new insights into the physics of the FQHE.

In this manuscript, we attempt to construct an exact interaction Hamiltonian for the LLL projected Jain composite-fermion (CF) wave functions given by 
\begin{equation}
\Psi_{n\over mn+1}=P_{\rm LLL} \Phi_n(z_1,z_2\dots) \prod_{i<j=1}^N(z_i-z_j)^m
\end{equation}
where $\Phi_n$ is the wave function of a fermionic integral quantum Hall effect (IQHE) state with $n$ Landau levels fully occupied and $P_{\rm LLL}$ is the LLL projection.
The right hand side is interpreted as $n$ filled ``$\Lambda$ levels," which are CF levels analogous to the electronic LLs. To this end,
we first study the behavior of these wave functions when a cluster of ${\cal N}$ particles is brought close together.
We find that the wave function in general vanishes faster
(\ie with a higher power of the distance )
than what is required by the Pauli principle in the case of fermionic FQHE states.
This allows us to identify missing configurations of the $n/(mn +1)$ states,
in analogy to the missing angular momentum pairs in the Laughlin $1/m$ state. 

The missing configurations for $n/(mn +1)$ states are consistent with the information encoded in the ``root partitions"\cite{Bernevig08}, or the ``DNA" \cite{Bandyopadhyay18}, of these states,
identified previously by Regnault, Bernevig and Haldane\cite{Regnault09}. 
Restricting to a Hilbert space consistent with the root partition (\ie retaining only the sub-dominant configurations that can be obtained by squeezing the root partition)
is equivalent to eliminating certain multiparticle configurations.
As noted in Ref.~\onlinecite{Regnault09},
restricting to a Hilbert space consistent with the root partition produces, for the 
$n/(mn +1)$ states with $n\neq 1$, several translationally invariant states (\ie states with total orbital angular momentum $L=0$ in the spherical geometry). This is in contrast to the examples of Laughlin, 
Read-Rezayi\cite{Read99}, and the Pfaffian states\cite{Moore91},
which are uniquely determined by the root partition function, \ie are the only $L=0$ states in the relevant Hilbert space. {\color{black}These states, unlike the CF states considered in our work, have periodic root partitions; FQHE states and their excitations characterized by such periodic root partitions have been classified in Refs.~\cite{PhysRevB.77.235108,PhysRevB.79.195132}.

{\color{black}Part of the objective of this work is to consider multi-particle interactions that 
go beyond constraints imposed by the root partition.} To avoid the Pauli zeros that occur in the fermionic wave functions and to focus on the correlation zeros alone, we will consider the bosonic state at $\nu=n/(n+1)$,
for which all of the zeros result from correlations; generalization to fermions by including the Pauli zeros is in principle straightforward.
These states represent the physics in which bosons capture a single vortex each to turn into composite fermions.
These wave functions have been found to be valid for bosons in a magnetic field interacting with a contact interaction \cite{Regnault03,Chang05b}.
We consider in detail the example of the bosonic CF state at $\nu=2/3$, which is the simplest state with non-trivial zeros.
There is no interaction in the two particle channel, as the state does not vanish when two bosons are coincident.
In the three particle channel the relative angular momentum ${\cal L}=0$ is absent
(\ie the CF wave function vanishes when three particles are coincident),
so we impose an energetic penalty for that configuration in our Hamiltonian.
In the 4-particle channel,
configurations with angular momenta up to (and including) ${\cal L}=5$ are absent in the wave function.
Including terms in the Hamiltonian that impose a penalty on those configurations still does not single out the CF state.
(We note that these constraints are not all independent.)
An inspection of 4 particle configurations in the ${\cal L}=6$ sector reveals further non-trivial correlations in the CF wave function,
which is the primary focus of our work.
There are three possible states of four bosons in the ${\cal L}=6$ sector, which are labelled $T_6$, $T_{42}$ and $T_{33}$. 
{These states differ (see Sec.~\ref{sec:Tmudefs} for details) in how the 4-particle wave function vanishes as a subset of the particles are brought into close proximity.
The wave function $T_6$ does not vanish when 2 or 3 particles are coincident,
but vanishes as 6 powers of the distance as the 4th particle approaches.
$T_{42}$ ($T_{33}$) does not vanish when 2 particles are coincident,
but the third particle brings 2 (3) zeros and the fourth 4 (3). $T_{66}$ is strictly absent in the Jain CF wave function.
While $T_{42}$ and $T_{33}$ are both present, they occur in a specific linear combination.}
We therefore consider a Hamiltonian that imposes an energetic penalty on the specific states\cite{Simon07} $T_6$ and $T_{42}+\lambda T_{33}$,
(in addition to the three particle channel with relative angular momentum $\mathcal{L}=0$)
and search for a zero energy solution as a function of $\lambda$,
{with the hope that the CF wave function will occur as the exact zero energy solution for some value of $\lambda$.} 

For $\lambda=0$, the so-called Gaffnian state\cite{Simon07b} is obtained as the exact and unique zero energy solution.
The Gaffnian state has a high overlap with the $\nu=\nicefrac{2}{3}$ CF state, but the two are topologically distinct, as can be ascertained by comparing their root partitions,
their entanglement spectra\cite{Regnault09} and the counting of excitations\cite{Toke09}. 

The CF state is not obtained as the exact and unique zero energy state for this model for any choice of $\lambda$. 
This indicates that no Hamiltonian up to (and including) 4-particle local interactions produces the CF state as an exact and unique zero energy ground state.
The reason is that in the CF wave function the constraint in question is satisfied for a single 4-particle cluster,
but it is not satisfied for all 4-particle clusters simultaneously. We speculate that no local interaction will produce the non-Laughlin LLL-projected CF wave functions as unique zero energy ground states. ({\color{black}We note that parent Hamiltonians can sometimes be approximated in finite systems.\cite{PhysRevB.98.081113}})

Surprisingly, we do find a zero energy state $\lambda \approx -0.18$ for systems with up to $N_b=12$ particles,
which we believe represents a zero energy state in the thermodynamic limit. {We refer to this state as the $\lambda$-state.
It is tempting to suspect that the $\lambda$-state is topologically equivalent to the CF state, and we perform tests to address this issue.
The following results suggest that the $\lambda$-state and the CF state are topologically equivalent: The two states have a high overlap ($0.98,0.97$ for systems with $\Nb=8,10$ bosons) and the same DNA,
and their neutral excitations and quasiparticles occur at the same quantum numbers. { \color{black} Constraints imposed by the root partition of the ground state on the generalized Pauli principles satisfied by the QH states suggest that the quasihole counting should resemble that of the  CF state.\cite{Seidel2010} However, the counting of low energy quasihole excitations do not match due to a vanishingly small gap above the QH states predicted expected from the CF state.}

We are therefore not able to make a definitive statement on the topological equivalence of the $\lambda$-sate and the CF state.} A comparison of the entanglement spectra would be useful in this regard\cite{Li08,Regnault09,Sterdyniak11,Sterdyniak12,Dubail12,Rodriguez12,Rodriguez13,Estienne15,Davenport15},
but previous work has shown \cite{Rodriguez12,Rodriguez13,Davenport15} that very large systems are required to obtain meaningful information from the entanglement spectra of the CF states; in the absence of an explicit wave function for the $\lambda$-state, such large systems are not accessible to us at the present.   We note that the $\lambda$-state is distinct from the Gaffnian state as the two have different root partitions. 

{\color{black}As pointed out in Refs.~ \onlinecite{green2002strongly,Simon07b}, many of the clustering rules produce states that may be gapless in the thermodynamic limit, such as the Haffnian or the Gaffnian state~\cite{Read09,PhysRevB.83.241302}. We find for the $\lambda$-state a rapid proliferation of the number of low energy ``quasihole states" upon addition of fluxes, potentially indicating a critical gapless state in the thermodynamic limit.} 

One may ask if there is a connection of the above results with the conformal field theory (CFT) construction of the FQHE wave functions.
The LLL projected CF-states can be written as correlation functions in a CFT approach\cite{Hansson07,Hansson07a,Kvorning13,Hansson17}.
A fundamental aspect of this construction is that the states corresponding to $n\geq 2$ filled $\Lambda$-levels necessarily require inserting a mix of both primary and descendent fields into the CFT correlator.
This is to be contrasted with the earlier constructions of the Laughlin, Pfaffian and Gaffnian wave functions\cite{Moore91,Simon07b} that require insertions of only primary fields. We find it tempting to speculate, without proof, that a construction of local exact parent Hamiltonians is possible only for wave functions that are generated with nothing but primary fields. 

We also mention that Chen {\em et al.} \cite{ChenLi2017} have argued that the generalized Pauli principle, rather than the root partition, determines the universal properties of a FQHE state. Considering the edge mode counting for the unprojected Jain wave function at $\nu=2/5$ they conjecture that a parent Hamiltonian does not exist that correctly produces the edge modes for the projected 2/5 wave function as zero modes.

The plan of the paper is as follows. The different symmetric polynomials used in the analysis of the lowest Landau level projected wave functions are introduced in section \ref{sec:symmetricpolynomials}. 
In section \ref{sec:short_distance} we discuss the short distance properties of the various symmetric polynomials when multiple particles are brought close together. Next,
in section \ref{sec:short_dist_CF} we focus on the discussion of short distance behaviour on the CF states.
The consequence of missing symmetric polynomials for the parent Hamiltonians is discussed in section \ref{sec:sym_pol_and_ham}.
Sections \ref{sec:Num_costruction} and \ref{sec:N_part_constraint} are devoted to the numerical implementation and search for a zero energy state,
followed by section \ref{sec:excitations} where the excitation spectrum of the earlier mentioned $\lambda$-state is explored. We end with conclusions. 

\section{Symmetric polynomials}
\label{sec:symmetricpolynomials}
In this section we define the notations for symmetric and translationally invariant polynomials that will be used in the rest of the paper.

\subsection{Monomial symmetric polynomials $P_\mu$}
In the symmetric gauge, the lowest Landau level single particle orbitals with angular momentum $m$ have the form  
\begin{equation}
\phi_{m}\propto e^{-\frac1{4}\left|z\right|^{2}}z^{m}\nonumber
\end{equation}
where $z=x+\imath y$.  
A many-body bosonic state can be expanded as 
\begin{equation}
\Psi\left(\mathbf{z}\right)=
e^{-\frac1{4}\sum_{i=1}^{N}\left|z_{i}\right|^{2}}
\sum_{\mathbf{m}=\left\{ m_1,\ldots,m_{N}\right\} }\alpha_{\mathbf{m}}P_{\mathbf{m}}\left(\mathbf{z}\right)
\label{eq:monomialsymmetricpolynomialexpansion}
\end{equation}
where the $\{P_{\bf m}\}$ forms a basis for the space of symmetric polynomials in $N$ variables $\mathbf{z}\equiv\{z_1,\dots,z_\NSymPol\}$ and $\alpha_{\mathbf{m}}$ are the coefficients.
In the following discussions,
we will omit the exponential factors $e^{-\frac1{4}\sum_{j=1}^{\NSymPol}\left|z_{j}\right|^{2}}$ and focus on the polynomial part of the wave functions.
These basis functions can be chosen to be the monomial symmetric functions
\begin{equation}
P_{\mathbf{m}}\left(\mathbf{z}\right)=\mathcal{S}
\left(\prod_{i=1}^{\NSymPol}z_i^{m_{i}}\right)
\end{equation}
where $\mathcal{S}$ is a symmetrizer. These are indexed by ordered integer partitions $\mathbf{m}\equiv\{ m_1,\dots m_N \}$ with $m_i\geq m_{i+1}$.
The basis functions represent states in which bosons occupy the single particle orbitals of angular momenta $m_1,m_2\dots,
m_N$ and therefore have a total angular momentum $\LAngMom$ of $\sum m_i$. 

To give a concrete example, for $\NSymPol=4$ bosons at $\LAngMom=6$,
there are nine polynomials labeled by  $m_1m_2m_3m_4\equiv 6000$, 5100, 4200, 4110, 3300, 3210, 3111, 2220 and $2211$.
The trailing zeros in the partitions will be suppressed when writing the polynomials. For instance,
an $\NSymPol=4$ particle function $P_{5100}=\mathcal{S}(z_1^5z_2^1z_3^0z_4^0)$ will be written as $P_{51}$.
The basis functions can be conveniently ordered based on the lexicographic ordering of the partitions that  label them.
An $\NSymPol$ particle basis $P_{\mu}$ is to the left of $P_{\mu^{\prime}}$,
if there exists a $1\leq k\leq \NSymPol$ such that $\mu_{j}=\mu_{j}^{\prime}$ for all $j>k$ and $\mu_{k}>\mu_{k}^{\prime}$.
For instance $P_{3210}$ is to the left of $P_{4110}$. 

\subsection{Translationally invariant symmetric polynomials}
In anticipation of the discussions to follow we now introduce the translation invariant symmetric polynomials.
These polynomials are invariant under rigid translations of the form $z_{i}\to z_{i}+a$ for all $i$. We will be working with two bases for these polynomials.
Following the notation of Refs [\onlinecite{Simon07},\onlinecite{Simon07a}] we will now construct basis functions representing $\NSymPol$ particle states with a well defined total relative angular momentum (total power of all the relative coordinates,
excluding the center of mass coordinate). The relative coordinates $\mathfrak{z}_j$ are given by 
\begin{equation}
\mathfrak{z}_{j}=z_{j}-\frac{\sum_{j=1}^{\NSymPol}z_{j}}{\NSymPol}\nonumber
\end{equation}
Elementary symmetric polynomials are defined in terms of the relative coordinates as follows 
\begin{eqnarray}
e_{k,\NSymPol} = \sum_{i_1<i_2\dots< i_k=1}^{\NSymPol}\mathfrak{z}_{i_1}\mathfrak{z}_{i_2} \dots\mathfrak{z}_{i_k}
\end{eqnarray}
and are indexed by the total relative angular momentum $k$ which can take values from $2,3\dots \NSymPol$. $e_{1,\NSymPol}$ is identically zero. For our purposes,
it is important to note that the total number of zeros of $e_{k,\NSymPol}$ when the function is viewed as a function of any of the coordinates $z_i$ is $k$.
The basis functions for the $\NSymPol$-particle translationally symmetric states $\phi_{\mu,\NSymPol}$ can be written in terms of the elementary symmetrical polynomials as
\begin{equation}
\phi_{\mu,\NSymPol}=\prod_{j=2}^{\NSymPol}\left[e_{j,\NSymPol}\right]^{\mu_{j}}\label{eq:philambda_trans_symm}
\end{equation}

The total angular momentum of the basis function is given by $\LAngMom=\sum_{j=2}^{\NSymPol}j\mu_{j}$.
The construction shows that for a given $\LAngMom$ and $\NSymPol$,
the number of linearly independent $\phi_{\mu,\NSymPol}$ is given by the number of partitions of the integer $\LAngMom$,
into integers between (and including) $2$ and $\NSymPol$ with repetitions allowed.
Each such non-distinct integer partition of $\LAngMom$ can be represented as $\mu \equiv 2^{\mu_2}3^{\mu_3}4^{\mu_4}\dots$,
where $\mu_i$ is the multiplicity (which can be $0,1,2\dots$) of $i=2,3\dots,\NSymPol$ in the partition.
Such a representation can then be mapped to $\phi_{\mu,\NSymPol}$ as defined in Eq.~\ref{eq:philambda_trans_symm}.
The basis functions for $\NSymPol=3$ and $4$ are listed in Table \ref{tab:trans_inv_34}.

\begin{table}[tb]
\begin{tabular}{|c|c|c|c|c|c|c|c|}
\hline 
$\LAngMom=$ & $0$ & $1$ & $2$ & $3$ & $4$ & $5$ & $6$\tabularnewline
\hline 
\hline 
$\NSymPol=3$ & $1$ & - & $e_{2,3}$ & $e_{3,3}$ & $\left[e_{2,3}\right]^{2}$ & $e_{2,3}e_{3,3}$ & $\left[e_{3,3}\right]^{2}$; $\left[e_{2,3}\right]^{3}$\tabularnewline
\hline 
$\NSymPol=4$ & $1$ & - & $e_{2,4}$ & $e_{3,4}$ & $\left[e_{2,4}\right]^{2}$; $e_{4,4}$ & $e_{2,4}e_{3,4}$ & $\left[e_{3,4}\right]^{2}$; $\left[e_{2,4}\right]^{3}$; $e_{2,4}e_{4,4}$\tabularnewline
\hline 
\end{tabular}
\caption{Translationally invariant states for $\NSymPol=3$ and $4$\label{tab:trans_inv_34}.}
\end{table}

The number of zeros of each coordinate $z_i$ in the state $\phi_{\mu,\NSymPol}$ is also $\LAngMom$. However,
if there is more than one basis function at a given $\LAngMom$,
then it is possible to combine them in a way as to make the largest power of $z_{i}$ less than $\LAngMom$. For example, for $\NSymPol=4$
particles at $\LAngMom=6$ there are three basis functions (Table \ref{tab:trans_inv_34}),
and it is possible to combine them so that the largest power of $z_1$ is only 3.

\begin{table*}[tb]
\begin{centering}
\begin{tabular}{|c|c|c|c|c|c|c|c|c|c|}
\hline 
 & $P_{6}$ & $P_{51}$ & $P_{42}$ & $P_{411}$ & $P_{33}$ & $P_{321}$ & $P_{3111}$ & $P_{222}$ & $P_{2211}$\tabularnewline
\hline 
\hline 
$T_{6}$ & $\frac{\sqrt{183}}{32}$ & $-\frac{\sqrt{\frac{183}{2}}}{16}$
& $\frac{23\sqrt{\frac{15}{61}}}{32}$ & $\frac{19\sqrt{\frac{15}{122}}}{16}$
& $-\frac{13}{16}\sqrt{\frac{5}{122}}$ & $-\frac{5}{8}\sqrt{\frac{15}{122}}$
& $-\frac{9}{8}\sqrt{\frac{5}{122}}$ & $\frac{7\sqrt{\frac{15}{122}}}{16}$
& $\frac{9\sqrt{\frac{5}{122}}}{16}$\tabularnewline
\hline 
$T_{42}$ &  &  & $\sqrt{\frac{43}{122}}$ & $-\frac{\sqrt{\frac{43}{61}}}{2}$
& $-\frac{33}{2}\sqrt{\frac{3}{2623}}$ & $\frac{13}{2\sqrt{2623}}$
& $10\sqrt{\frac{3}{2623}}$ & $-\frac{3}{2\sqrt{2623}}$
& $-5\sqrt{\frac{3}{2623}}$\tabularnewline
\hline 
$T_{33}$ &  &  &  &  & $\sqrt{\frac{3}{43}}$ & $-\frac{3}{\sqrt{43}}$ & $2\sqrt{\frac{3}{43}}$ & $\frac{4}{\sqrt{43}}$ & $-\sqrt{\frac{3}{43}}$\tabularnewline
\hline 
\end{tabular}
\end{centering}
\caption{Expansion of translationally invariant states $T_\mu$ for $\NSymPol=4$ in terms of monomial symmetric polynomials $P_\mu$\label{tab:decomp_4}.}
\end{table*}

\subsection{Translation invariant symmetric polynomials $T_\mu$}
\label{sec:Tmudefs}
Due to the possibility of creating linear combinations of $\phi_{\mu,\NSymPol}$ where the highest power of $z_1$ is lower than $\LAngMom$,
it is useful to introduce a different type of polynomial $T_{\mu}$.
These polynomials are orthonormal linear combinations of the $\phi_{\mu,\NSymPol}$, and are therefore still translationally invariant.
The specific linear combinations are constructed in such a way that they have different vanishing properties (described in the next section) as well as different highest powers of the $z_i$ in them.

To make the idea precise, we choose the example of $\NSymPol=4$, and $\LAngMom=6$.
As seen in table \ref{tab:trans_inv_34} there are three linearly independent polynomials $\phi_{222,4}= [e_{2,4}]^{3}$,
$\phi_{33,4} = [e_{3,4}]^{2}$ and $\phi_{42,4} = e_{2,4}e_{4,4}$.
These will all contain the monomial $P_{6}$,
but any one of the states can be used to eliminate through linear combinations
the $P_6$ terms in the remaining two.
Since the monomial symmetric polynomials $P_\mu$ form an orthonormal basis,
this implies that there is a two dimensional subspace of the space spanned by $\{\phi_{\mu,\NSymPol}\}$ which is orthogonal to $P_6$.
We define the unique state orthogonal to this two dimensional space as the state $T_6$.
\ie\, $T_6$ is the orthogonal projection of $P_6$ in   $\{\phi_{\mu,\NSymPol}\}$.

We can now iterate this step by restricting to the (two dimensional in the example) space orthogonal to $T_{6}$ and within this space sequentially identify states $T_\mu$ that are the projections of $P_\mu$'s.
The states obtained this way depends on the ordering of the $\mu$ sequence. The reverse lexicographic ordering provides a natural ordering for the purpose.
In the case of $\NSymPol=4$ and $\LAngMom=6$, the sequence is  $6\to51\to42\to411\to33$

This construction defines the translation invariant symmetric polynomials $T_{42}$ and $T_{33}$ in addition to $T_6$ as shown in Table \ref{tab:decomp_4}. By construction these states are orthogonal.
Note that not all $T_\mu$'s occur - for instance there are no states of the form $T_{51}$ or $T_{411}$.
Identification of states $T_{\mu}$ in the translationally invariant subspace can be achieved for general $\LAngMom$ and $\NSymPol$. For the case of $\NSymPol=4$ and $\LAngMom=4$,
there is a two dimensional space of translationally invariant states spanned by $T_{4}$ and $T_{22}$.
The sequence of steps in the construction of $T_{\mu}$ can be used to identify corresponding states in the spherical geometry also as will be described in Sec \ref{sec:TmuOnSphere}.
In the following section we will discuss the short distance properties of $T_\mu$ states.

\section{Short distance behaviour}\label{sec:short_distance}

The $T_{\mu}$ polynomials can be characterized by the manner in which the polynomial vanishes when $\NSDGroup<\NSDTotal$ particles are brought close together.
Due to the holomorphic nature of the wave function,
this is closely related to the number of vortices attached to any cluster of $\NSDGroup-1$ particles.
In this section we make these notions more precise.

\begin{figure}[h!]
\includegraphics[width=\columnwidth]{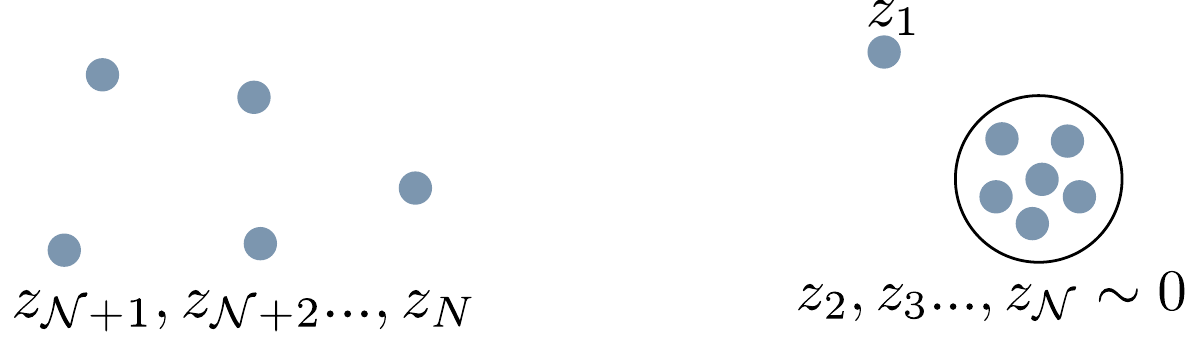}
\caption{Schematic illustration for the clustering of configuration of the $N$ particles for a calculation of the number of zeros $\CalF(\mathcal{N})$ (see text). \label{fig:schematicClustering}}
\end{figure}
Consider the $\NSDTotal$ particle wave function as a polynomial $f(z_1)$ in the coordinate $z_1$ while $\NSDGroup-1$ are fixed in the vicinity of a point say $\eta=0$ (Fig.
\ref{fig:schematicClustering}). The remaining particles are far from this point,
\ie\,$0<|z_{i}-\eta|\sim\epsilon\ll 1$ at some not fine-tuned locations for $i=2,3,\dots,\NSDGroup$ and $|z_{i}-\eta|\gg1$ for $i=\NSDGroup+1,\dots \NSDTotal$.
The number of zeros of the polynomial $f(z_1)$ in the vicinity of $\eta$ is equal to the number of vortices that the particle $z_1$ sees in the $\NSDGroup-1$ particle cluster,
and we will call it the ``zeros attached to a $\NSDGroup$ particle cluster" and denote it as $\CalF(\NSDGroup)$.
In general the number of attached zeros are independent of the precise locations of the $\NSDTotal-\NSDGroup$ distant particles.

As an example we look at the $\NSDTotal=4$ particle monomial symmetric polynomial state $P_{42}$ which is $S(z_1^{4}z_{2}^{2})$. For $\NSDGroup=1$ and $\NSDGroup=2$,
the polynomial does not vanish when $z_{1,2}$ approaches $0$ since there will be terms to the type $z_{3}^{4}z_{4}^{2}$ that are nonzero.
However for $\NSDGroup=3$ the situation in different. When the particles $z_{3},z_4$ are fixed in the $\epsilon$ vicinity of $0$,
the wave function is $z_1^2 (z_2^2 z_1^4 + z_2^4)+\mathcal{O}(\epsilon^4)$ which as a function of $z_1$ has two zeros near $\epsilon$.
The precise location of these two zeros depends on the location of $z_2$.
The remaining two zeros of this polynomial appear in the vicinity of $z_2$.
In a similar manner for $\NSDGroup=4$, $z_1$ sees four zeros in the vicinity of $z_2,z_3,z_4\sim \epsilon \to 0$ as the wave function in this case behaves as $P_{42}\sim z_1^4 \epsilon^2 + \mathcal{O}(\epsilon^4)$.

Similar calculation for $P_{33}$ shows that it does not vanish for $\NSDGroup=1,2$ but for $\NSDGroup=3$ and $4$ the polynomial behaves as
$P_{33}\sim z_1^{3}z_{2}^{3}+\mathcal{O}(\epsilon^3)$ and $P_{33}\sim z_1^{3}\epsilon^{3} + \mathcal{O}(\epsilon^6)$ respectively,
and therefore both are associated with $3$ zeros. 

In general it can be shown that the number of zeros attached to $\NSDGroup$ particles in the wave function $P_\mu$ of $\NSDTotal$ bosons is equal to $\mu_\NSDGroup$
($\NSDGroup^{\rm th}$ member of the partition $\mu$)
when $\mu = \mu_\NSDTotal,\mu_{\NSDTotal-1}\dots \mu_1$ is written as a length $\NSDTotal$ partition
(padded with 0s as needed) sorted in the order
$\mu_{i+1} \leq  \mu_i$.
For example,
the number of zeros attached to $\NSDGroup=1,2,3,4$ particles in $P_{42}\equiv P_{4200}$ is $0,0,2,4$ respectively.
Similar characterization using attached zeros can be applied also to $T_{\mu}$.
The number of attached zeros can be inferred from the expansion in monomial symmetric polynomials $P_\mu$.
For this we use the folowing results regarding the zeros for a linear combination of different $P_{\mu}$s.

Consider the zeros of the sum of two polynomials  $P_{\ldots\mu_{3}\mu_{2}\mu_1}+
P_{\ldots\mu_{3}^{\prime}\mu_{2}^{\prime}\mu_1^{\prime}}$.
It can be shown that for $\NSDGroup>1$ the number of zeros $\CalF(\NSDGroup)$ is given by 
\begin{equation}
\CalF(\NSDGroup)=\begin{cases}
\mu_{\NSDGroup} & \text{if }\sum_{j<\NSDGroup}\left(\mu_{j}-\mu_{j}^{\prime}\right)<0\\
\mu_{\NSDGroup}^{\prime} & \text{if }\sum_{j<\NSDGroup}\left(\mu_{j}-\mu_{j}^{\prime}\right)>0\\
\min\left(\mu_{\NSDGroup},\mu_{\NSDGroup}^{\prime}\right) & \text{if }\sum_{j<\NSDGroup}\left(\mu_{j}-\mu_{j}^{\prime}\right)=0
\end{cases}
\end{equation}
A partition $\mu$ is said to dominate another partition $\mu^\prime$ if $\sum_{i<k} \mu_i $ is less than $\sum_{{i<k}} \mu^\prime_i$ for every $k=1,2\dots \NSDTotal$.
If a linear combination of partitions \ie\,$\sum_{\mu \in X}c_\mu P_\mu$,
$c_\mu\neq 0$ is such that $X$ contains a partition $\mu_0$ that dominates all other partitions in the set $X$,
then the counting of the zeros of the set will be determined by $P_{\mu_0}$.

For the $T_{\mu}$ defined above, this result implies that the structure of their zeros is the same as those of $P_{6}$,
$P_{42}$ and $P_{33}$ respectively. This is because the partitions $6000$,
$4200$ and $3300$ dominate all the other partitions occurring in the respective polynomials.
The vanishing properties of $T_\mu$ are summarized in Table \ref{tab:zeros_of_T}.

We note that when linear combinations do not contain a dominant partition,
vanishing properties of the combination is not given by any single monomial symmetric polynomial $P_{\mu}$.
For instance, in the case of $\NSDTotal=4$,
$\LAngMom=6$ the state $P_{3111}+P_{2220}$ has the vanishing sequence (zeros attached to $\NSDGroup=1,2,3,4$ particles): $0\to2\to1\to3$,
which means that one of the two zeros that where bound to two particles will pull free when a third particle is brought in.
Note that this sequence is distinct from that of $P_{3111}$ or $P_{2220}$ and that neither $3111$ nor $2220$ is dominant over the other.

The reader might wonder whether it is always possible to write translationally invariant state space spanned by basis states each one dominated by a single partition.
This question was answered with a definitive `No' in Ref.~\onlinecite{Liptrap2008},
where they construct an explicit counter-example for 4 particles at total angular momentum 14.
Nevertheless, it seems that the CF states do not suffer from this exception, and are always dominated by a single root partition.

\begin{table}[tb]
\begin{tabular}{|c|c|c|c|c|}
\hline  & $\NSDGroup=1$ & $\NSDGroup=2$ & $\NSDGroup=3$ & $\NSDGroup=4$  \\ \hline
$T_6$ & 0 & 0 & 0 & 6 \\ \hline
$T_{42}$ & 0 & 0 & 2 & 4 \\ \hline
$T_{33}$ & 0 & 0 & 3 & 3 \\ \hline
$P_{3111}+P_{2220}$ & 0 & 2 & 1 & 3 \\ \hline
\end{tabular}
\caption{Number of zeros $\CalF$ attached to an $\NSDGroup$ particle cluster for the functions $T_6$, $T_{42}$, and $T_{33}$.}
\label{tab:zeros_of_T}
\end{table}

\section{Short distance behaviour of CF States}\label{sec:short_dist_CF}

In this section we will explore the short distance behaviour of the composite fermion wave function for bosons at filling fraction $2/3$.
For earlier studies of the root partitions of CF states see Refs.~\onlinecite{Bernevig08,Regnault09,Rodriguez12b}.

For the purposes of this section, we will represent the state with $N$ bosons or fermions occupying the single particle orbitals $\mathbf{m}=(m_1,m_2\dots m_N)$ as
\begin{equation}
  ({\mathbf{m}})_+ = S\left[\prod_{j}z_{j}^{m_{j}}\right],
  \quad\quad
({\mathbf{m}})_- = A\left[\prod_{j}z_{j}^{m_{j}}\right],\label{eq:m_pol}
\end{equation}
respectively.
Here $S$ and $A$ represent the  symmetrization and the antisymmetrization operation.
The bosonic state $({\mathbf{m}})_+$ is same as the monomial symmetric polynomials.
The fermionic state is the determinant of a matrix with matrix elements $M_{ij} = z_i^{m_j}$.
Permutation of indices by $\sigma$ leads to equivalent polynomials with $(\mathbf{m})_+=(\sigma(\mathbf{m}))_+$
and $(\mathbf{m})_-=(-1)^\sigma (\sigma(\mathbf{m})_-)$.
In the following discussion,
we will generally drop normalization factors and signs,
as we are mainly interested in the leading order behaviour when several particles are brought close together.

 In a similar fashion as \eqref{eq:m_pol}, one can construct polynomials of derivative operators as follows
\begin{equation}
  ({\mathbf{\bar{m}}})_+ = S\left[\prod_{j}\partial_{{j}}^{m_{j}}\right],
  \quad\quad
({\mathbf{\bar{m}}})_- = A\left[\prod_{j}\partial_{{j}}^{m_{j}}\right],
\end{equation}
where $\partial_{j}\equiv\frac{\partial}{\partial z_{j}}$. 

Multiplication of symmetric polynomials define an algebra for $N$-particle bosonic functions $({\mathbf{m}})_+$ and the derivative operation on the polynomials define the linear action of $({\mathbf{\bar{m}}})_+$ on $(\mathbf{m})_+$.
When two $N$-particle symmetric polynomials $({\mathbf{m}})_+$ and $({\mathbf{n}})_+$ are multiplied together the resulting polynomial is a linear combination
\begin{eqnarray*}
(\mathbf{m})_+ (\mathbf{n})_+ = \sum_{\sigma \in S_N} (\mathbf{m} + \sigma(\mathbf{n}))_+,
\end{eqnarray*}
where $S_N$ is the set of all permutations of $N$ objects.
The action of $(\mathbf{\bar{m}})_+$ on $(\mathbf{n})_+$ is given by
\begin{eqnarray*}
  \left(\mathbf{\bar{m}}\right)_{+}\left(\mathbf{n}\right)_{+}=
  \sum_{\sigma}\left[\prod_{j}\frac{n_{j}!}{\left(n_{j}-m_{\sigma\left(j\right)}\right)!}\right]\left(\mathbf{n}
  -\sigma\left(\mathbf{m}\right)\right)_{+}
\end{eqnarray*}
where it is assumed that $(\mathbf{m})_+=0$ if any of the $m_i$ are negative.
It is also useful to define a representation for the combination of derivatives and polynomials as 
\begin{eqnarray*}
\left(\mathbf{\bar{m}l}\right)_{+}=\left(\bar{m}_1l_1,\bar{m}_{2}l_{2},\ldots,\bar{m}_{N}l_{M}\right)_{+}=\mathcal{S}\left[\prod_{j}\partial_{j}^{m_{j}}z_{j}^{l_{j}}\right].
\end{eqnarray*}
The action of $(\mathbf{\bar{m}l})_+$ on $(\mathbf{n})_+$ gives
\begin{eqnarray*}
  \left(\mathbf{\bar{m}l}\right)\left(\mathbf{n}\right)=
  \sum_{\sigma}\left[\prod_{j}\frac{\left(n_{j}+l_{\sigma\left(j\right)}\right)!}{\left(n_{j}+l_{\sigma\left(j\right)}-m_{\sigma\left(j\right)}\right)!}\right]
  \left(\mathbf{n}+\sigma(\mathbf{l}-\mathbf{m})\right).
\end{eqnarray*}

The fermionic state $\prod_{i<k=1}^N (z_i-z_k)$ which describes the integer quantum Hall effect at filling fraction $1$ is identical to $(0,1,2,\dots N-1)_-$.
General Laughlin type states $\prod_{i<k=1}^N (z_i-z_k)^q$ which occurs at filling fraction $1/q$ cannot be represented by a single term like this.
However, the state can be written as a linear combination of polynomials $(\mathbf{m})_{(-)^q}$,
all of which are related to a single monomial symmetric polynomial $\rho_q$,
given below, by the so called squeezing rules \cite{Bernevig08}.
\begin{equation}
\rho_{q}=\left(0,1q,2q,3q,\ldots,q\left(N-1\right)\right)_{\left(-\right)^{q}}.
\end{equation}
The set of partitions that can be obtained by squeezing rules satisfy the property that they are dominated (as defined in the Sec. \ref{sec:short_distance}) by the partion $\rho_{q}$.

\subsection{CF states}

CF states for bosons (fermions) on the plane for $\frac{n}{np+1}$, with $p$ odd (even), for $N=nc$ particles
are built by first filling $n$ $\Lambda$-levels with $c$ particles in each level
and then multiplying with the Jastrow factor 
$\psi_{\frac1{p}}=\prod_{i<j=1}^N (z_i-z_j)^{p}$. The state is the projected to the LLL.
For $n=1$, which gives the Laughlin type states at filling fraction $\frac1{p+1}$,
the projection is superfluous and we can directly write this state as
\begin{eqnarray*}
\phi & = & \begin{vmatrix}1 & z_{N} & z_{N}^{2} & \cdots & z_{N}^{N-1}\\
\vdots & \vdots & \vdots &  & \vdots\\
1 & z_{3} & z_{3}^{2} & \cdots & z_{3}^{N-1}\\
1 & z_{2} & z_{2}^{2} & \cdots & z_{2}^{N-1}\\
1 & z_1 & z_1^{2} & \cdots & z_1^{N-1}
\end{vmatrix} \times  \psi_{\frac1{p}}.
\end{eqnarray*}
In the notation developed in the previous section, this state can be written as 
\begin{equation}
\phi = \left(0,1,2,3,\ldots,N-1\right)_{-}\times \psi_{\frac1{p}}.
\end{equation}

For $n=2$, which gives the states in the sequence $\frac{2}{2p+1}$, we have
\begin{eqnarray*}
\phi & = & P_{\rm LLL} \begin{vmatrix} &  & \vdots &  &  &  & \vdots\\
1 & z_1 & \cdots & z_1^{c-1} & \bar{z}_1 & \bar{z}_1z_1 & \cdots & \bar{z}_1z_1^{c-1}
\end{vmatrix}\psi_{\frac1{p}}.
\end{eqnarray*}
The projection can be performed exactly by the replacement $\bar z\to \partial_z$ (where we omit scale factors),
and the understanding that the derivatives act all the way from the left.
The determinant is then written as 
\begin{eqnarray*}
\phi & = & \begin{vmatrix} &  & \vdots &  &  &  & \vdots\\
1 & z_1 & \cdots & z_1^{c-1} & \partial_1 & \partial_1z_1 & \cdots & \partial_1z_1^{c-1}
\end{vmatrix}\psi_{\frac1{p}}
\end{eqnarray*}
which in the notation previously developed translates to
\begin{equation}
\left(0,1,\ldots,c-1,\bar1,\bar11,\ldots,\bar1\left(c-1\right)\right)_{-} \times \psi_{\frac1{p}}.
\end{equation}
The case of generic $n$ proceeds 
similarly as for the two previous examples, but now derivatives up to $\partial^{n-1}$ are included.

\subsection{Vanishing properties of CF states}
As described above, the projected composite fermion states can be written in terms of the action of $(\mathbf{\bar{m}l})_-$ on $\psi_{\frac1{p}}$.
This results in a large sum when written in terms of monomial symmetric polynomials.
However, like the Laughlin state $\psi_{\frac1{q}}$,
it still remains true that the overall wave function is a linear combination of polynomials $(\mathbf{m})$ with a single dominant partition $\rho$.\cite{Regnault09} 
As discussed in the previous section,
the vanishing properties of such wave functions are determined entirely by the dominant partition which can often be identified without full expansion.
In the following sections,
we use this to understand the vanishing properties of different composite fermion states.

\subsubsection{CF state for bosons at  $\nu=\frac{2}{3}$}

The wave function for the bosonic state at filling fraction $2/3$ is given by
\begin{multline}
\psi_{\frac{2}{3}}=(0,1,2,...., \bar1,\bar11,\bar12,\bar13\dots)_- (0,1,2,3,4\dots)_-\\
\equiv (\bar1,\bar11,0,\bar12,1,\bar13,2,\dots)_-  (0,1,2,3,4\dots)_-
\end{multline}
where the equivalence up to sign of $(\mathbf{\bar{m}l})_-$ under permutations of $\mathbf{\bar{m}l}$ has been used in the second line to arrange it in increasing order of $-m_i+l_i$.
With such an ordering the dominant partition can be obtained from considering the action of the individual terms in $(\mathbf{\bar{m}l})$ on $(0,1,2,3\dots)$. This is illustrated in the following:
\begin{multline}
\left(\bar1,\bar11,0,\bar12,1,\bar13,2,\bar14,3,\cdots\right)_{-}\left(0,1,2,3,4,5,\ldots\right)_{-}\\
= \left(\overbrace{0}^{0(0)},\overbrace{0}^{\bar1(1)},\overbrace{2}^{\bar11(2)},\overbrace{4}^{1(3)},\overbrace{5}^{\bar12(4)},
\overbrace{7}^{2(5)} \overbrace{8}^{\bar13(6)}\ldots\right)_{+}+\dots\nonumber
\end{multline}
The entries above the braces suggest one of the many possible ways in which the term below could have been produced.
Here the $+\cdots$ after the root partition stands for terms that are related
to the root partition via squeezing rules.

To understand the root partition obtained upon the action of $D=(\mathbf{\bar{m}l})$ on $(0,1,2,\dots)$ it is sufficient to know the sequence representing the order $-m_i+l_i$ of each term in $D$. In the case of $2/3$ state the order is given by 
\begin{equation}
(-1,0,0,1,1,2,2,3,3,4,4\dots)\nonumber
\end{equation}
The following expression shows how the order shown above of the terms in $D$ is related to the root partition.
\begin{equation}
\left(\overbrace{0}^{0(0)},\overbrace{0}^{-1(1)},\overbrace{2}^{0(2)},\overbrace{4,5}^{1(3,4)},\overbrace{7,8}^{2(5,6)}, \overbrace{10,11}^{3(7,8)} \ldots\right)_{+}\nonumber
\end{equation}
The entries above the braces written as $a(b,c)$ show that the parts below the braces result from the action of terms in $D$ of order $a$  on monomials of order $b$ and $c$. 

As discussed in the previous section, the number of attached zeros when $\NSDGroup$ particles are brought close togehter can be inferred from the root partition.
We thus find that $0,0,2,4,5,7\dots$ zeros are attached to clusters of $\NSDGroup=1,2,3,4,5,6\dots$ particles. Note that at large powers,
the sequence of the monomial powers is that of the Tao-Thouless-pattern on the torus\cite{Bergholtz08b},
showing that the root partition has the correct density at large angular momenta.
Let us also point out that the preceding discussion also applies to the case of the Jain-Kamilla (JK) projected CF wave functions,
since these only differ from the exactly projected CF wave functions in sub-dominant terms.
\subsubsection*{Fermionic state at $\nu=\frac{2}{5}$}

The vanishing properties of the fermionic wave function at $2/5$ are related to the vanishing properties of the $2/3$ state by $\CalF_{\frac{2}{5}}(\NSDGroup) = \CalF_{\frac{2}{3}}(\NSDGroup) + \NSDGroup-1$,
where the added term is the number of the Pauli zeros arising from the fermionic statistics between the $z_1$ and remaining particles in the cluster.
For a graphical illustration of the zeros attached to the $\NSDGroup$ particle cluster, see Fig. \ref{fig:2_5_zeros}.
The figure shows color plots of $\arg{\psi(\mathbf{z})}$ as a function of $z_1$ when the remaining particles are in fixed positions.

\begin{figure*}
\includegraphics[width=0.8\textwidth]{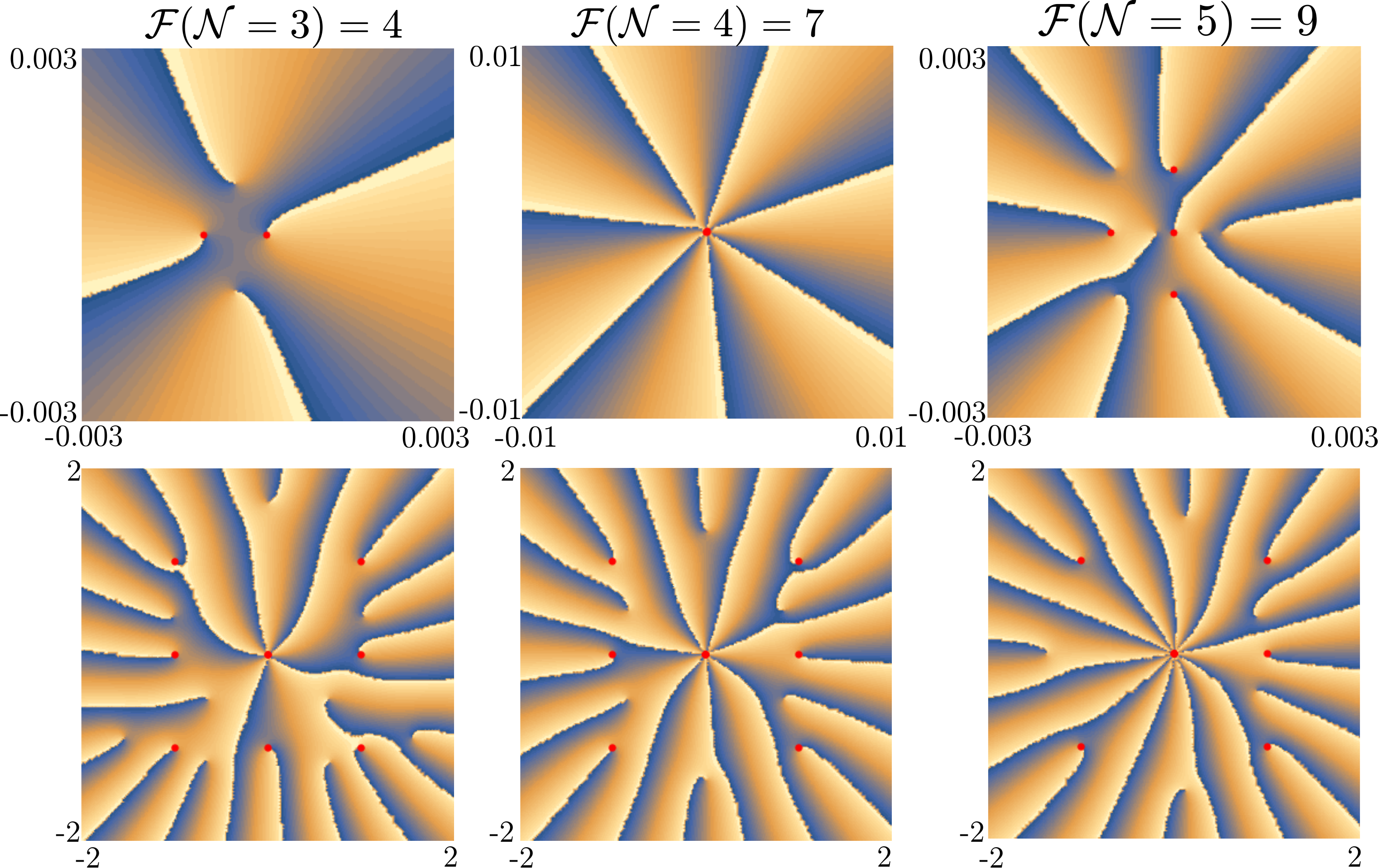}
\caption{
  Illustration of the vanishing properties of the fermionic $\nu=\frac{2}{5}$ state for $N=10$ particles.
  The plot shows the phase of the wave function, treated 
  as a function of one of the coordinates $z_1$ with the other $N-1=9$ particles held at fixed positions (red dots).
  Branch cuts represent jump in the phase from $0$ to $2\pi$.
  They terminate at the zeros of the wave function.
  (top) When $\NSDGroup=2$ (not shown), $3$ (left), $4$ (middle),
$5$ (right) particles are brought close to a common point $0$,
  the cluster attaches $\CalF(\mathcal{N})=1,4,7,9$ zeros respectively.
  The separations between the zeros decrease as inter-particle separations decrease.
  (bottom) At finite distance from the group of $\NSDGroup$ particles,
  there exist additional zeros.
  The figure shows $5$ (left), $7$ (middle) and $10$ (right) zeros (instead of $4,7,9$) near $0$.
  These additional zeros are closer to the group of $\NSDGroup$ particles than to any other particle but 
  these additional zeros stay at a finite distance from the remaining zeros as inter-particle distances in the cluster approach zero.
}\label{fig:2_5_zeros}
\end{figure*}

\subsubsection{Bosonic state at $\nu=\frac{n}{n+1}$ }

The vanishing properties of the $\nu=\frac{3}{4}$ state can also be derived in a similar manner.
Since the wave function has composite fermions occupying three $\Lambda$ levels,
the wave function representing the state can be written as $\psi_{3/4}=D (0,1,2,3\dots )_-$ where the $D$ operator is given by
\begin{multline}
D = (\bar{2},\bar{2}1,\bar{2}2\dots,\bar1,\bar11,\bar12\dots,0,1,2,\dots)\\
\equiv 
(\bar{2},\bar{2}1,\bar1,\bar{2}2,\bar11,0,\bar{2}3,\bar12,1,\dots)\nonumber.
\end{multline}
The order of terms in $D$ after sorting are
\begin{equation}
(-2,-1,-1,0,0,0,1,1,1,2,2,2,3,3,3\dots)
\end{equation}
which results in the root partition
\begin{equation}
\left(\overbrace{0}^{0(0)},\overbrace{0}^{-1(1)},\overbrace{0}^{-2(2)},\overbrace{2}^{-1(3)},\overbrace{4}^{0(4)},\overbrace{5}^{0(5)},
\overbrace{7,8,9}^{1(6,7,8)}, \overbrace{11,12,13}^{2(9,10,11)},\ldots\right)_{+}\nonumber.
\end{equation}
The entries above the braces indicate the order of the terms from $D$ and the order of the terms in the monomials which result in the parts of the root partition below the braces.
A similar calculation yields the root partition for the bosonic state at $\nu=4/5$ [containing 4 filled $\Lambda$-levels] to be 
\begin{equation}
(0,0,0,0,2,4,5,7,8,9,11,12,13,14,\dots).
\end{equation}

In the case of a general state in this sequence occurring at filling fraction $\frac{n}{n+1}$, the order of the terms in the derivative operator is 
\begin{equation}
  D=\left(-n+1,\overbrace{-n+2}^{\times2},\cdots,\overbrace{-1}^{\times n-1},\overbrace{0}^{\times n},\overbrace1^{\times n},\overbrace{2}^{\times n},\ldots\right),
  \nonumber
\end{equation}
where the entry $\times s$ above the curly braces says that there are $s$ successive copies of the entry below.

Acting with $D$ on $(0,1,2,3,\dots)$ gives 
\begin{equation}
{D}\psi_1=\left(\overbrace{0}^{\times n},\overbrace{2}^{\#1},\overbrace{4,5}^{\#2},\overbrace{7,8,9}^{\#3},\overbrace{11,12,13,14}^{\#4},\cdots\right)\label{eq:n/(2n+1)}
\end{equation}
The $\times n$ above the $0$ is to point out that there are $n$ copies of $0$.
The symbol $\# s$ points out that there are $s$ successive numbers below the braces.
The number of such successive terms increases until it reaches $n$,
beyond which the number of successive terms below the braces stays at $n$.
There is a gap of $2$ between entries below successive braces.
This pattern generalizes the root partitions found for the case of $2/3$, $3/4$ and $4/5$ states.

A generalization to $p>1$ states (for $\nu=\frac{n}{pn+1}$) proceeds in the same way as above,
with the exception that sometimes (for $p\geq 3$) it is necessary to consider also squeezing of the root partition of $\psi_{\frac1p}$ in order to obtain the correct vanishing properties.

\section{Missing Symmetric polynomials and projection Hamiltonians}\label{sec:sym_pol_and_ham}

\begin{figure*}
\includegraphics[width=0.9\textwidth]{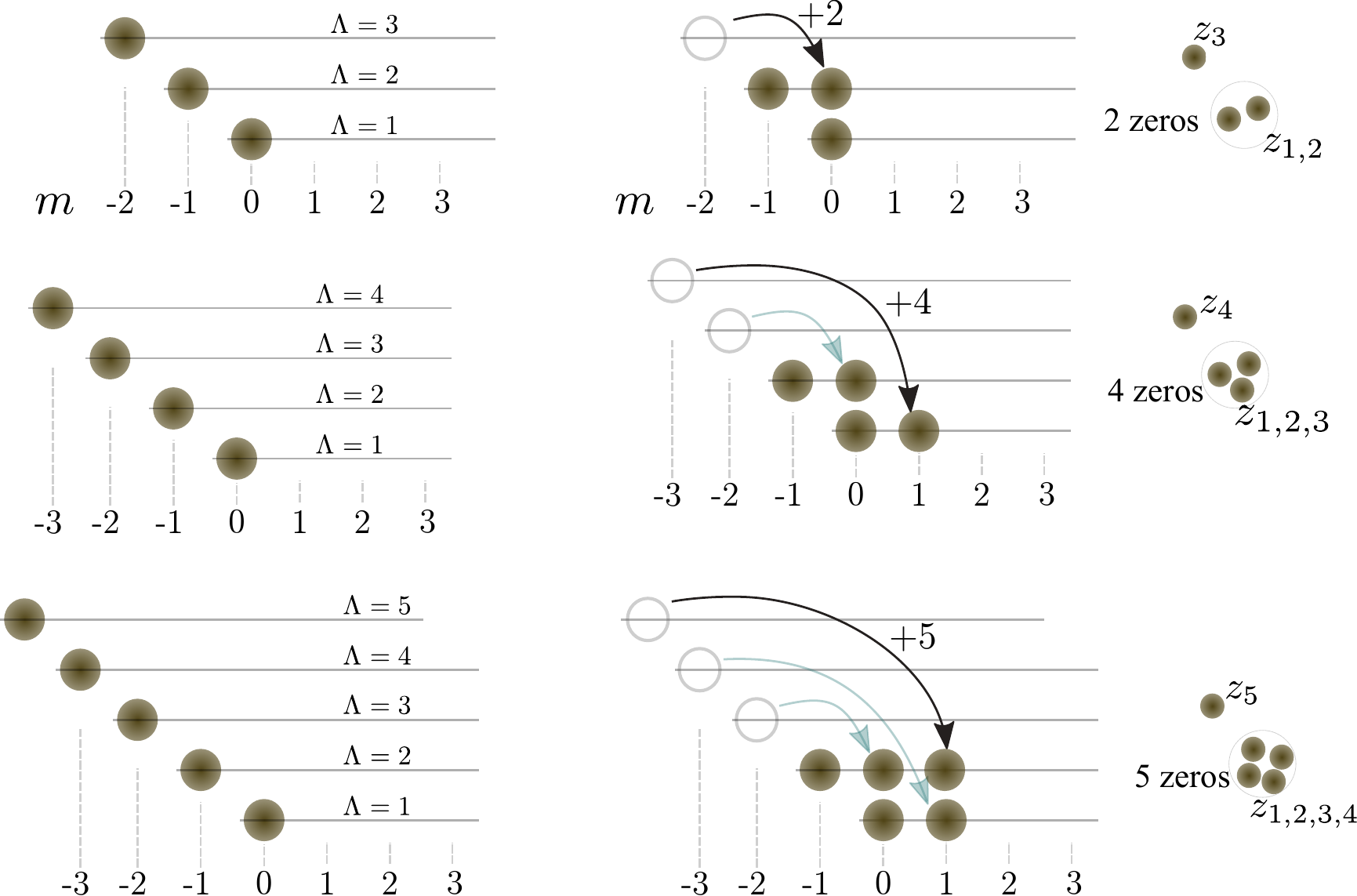}
\caption{	\label{fig:CFstructure}
  $\Lambda$-level description of the four-particle $p=1$ bosonic CF state with minimum total angular momentum (left) without any restriction ($\LAngMom=0$);
  (right) when only two $\Lambda$ levels are allowed for filling factor $\nu=2/3$ 
}
\end{figure*}

The analysis of the number of zeros attached to clusters in the previous section
implies that certain translationally invariant symmetric polynomials are missing in clusters of finite numbers of particles in the CF states.
These constraints on the clusters of particles can be used to construct projection operators for which the CF state forms a zero energy state.
We will confine the discussion to that of the bosonic $2/3$ state hereafter.

To illustrate the idea, we consider the case of $\NSDGroup=4$ particles, in the angular momentum sector $\LAngMom=6$.
there are three possible basis functions $T_6, T_{42}$,
and $T_{33}$ for translationally invariant states this sector.
The number of zeros attached to a $4$-particle cluster in these functions are $6,4$ and $3$ respectively.
However when any four particle cluster in a $\nu=2/3$ bosonic CF state is brought together,
the number of locked zeros is $4$ as can be inferred from the root partition of the CF state.
If the 4 particle state has any finite probability of being in the $T_6$ state,
the number of zeros attached to the $4$ particles will be $6$ ($P_6$ dominates $P_{42}$ ).
Thus no four particle cluster in the $2/3$ bosonic state can be in the state $T_6$.
Instead they will always be in a linear combination $T_{42}+\lambda T_{33}$.
The CF state should therefore be in the null space of the operator
\begin{equation}
\sum_{i<j<k<l=1}^N \mathcal{P}_{ijkl}\left[T_6\right]
\end{equation}
where $\mathcal{P}_{ijkl}\left[T_6\right]$ projects the state of the four particles $i,j,k,l$ into the state $T_6$.

For the case of $\NSDGroup=4$ particles, we will now consider the possibility of them being in a translationally invariant $\LAngMom=4$ channel. 
There are two translationally invariant symmetric polynomials $T_{4000}$ and $T_{2200}$ in this sector.
Neither of them or any linear combination can result in the vanishing sequence of the CF state $0,0,2,4$.
Consider instead the possibility of the four particles being in a linear combination with the $\LAngMom=6$ sector : $\varphi=aT_{4000}+bT_{2200}+cT_{4200}$.
Upon bringing $3$ particles close to each other,
this state will attach no zeros (instead of $2$ for the CF state) due to the $T_{4000}$ term. This implies that $T_{4000}$ is absent in the CF state.
Upon bringing $4$ particles close together in the state $\varphi$,
the state $T_{2200}$ leads to attaching of only $2$ zeros (instead of $4$ in the CF state).
This tells us that $T_{2200}$ is also not a possible state for the four particle clusters in of the CF state.
Together this implies that the CF state is in the null space of the operator
\begin{equation}
\sum_{i<j<k<l=1}^N \mathcal{P}_{ijkl}\left[\LLockZero=4\right]
\end{equation}
where $\mathcal{P}_{ijkl}\left[\LLockZero=4\right]$ is the projector into the space of relative angular momentum $\LLockZero=4$ state.

Similar considerations tell us that the CF state cannot have clusters of four particles in translationally invariant channels of angular momenta $\LAngMom=0,1,2,3,4$ and $5$.
For clusters of $\NSDGroup=3$ particles,
similar analysis tells us that translationally invariant $\LAngMom=0$ state is not present in the CF state.
Since the number of attached zeros is $0$ for $\NSDGroup=2$, there appears no constraints on the two particle clusters.

In summary, the analysis above tells us that the CF state should be in the null space of the projectors into the 
\begin{enumerate}
\item  4-particle channels of relative angular momenta $\LAngMom\leq 5$, and 
\item  4-particle channels in the state $T_6$.
\item  3-particle channels of relative angular momenta $\LAngMom=0$.
\end{enumerate}

These constraints are however not independent. To see this,
consider the nullspace of the projector on the $\LAngMom=0$ channel of $3$ particles.
Any three particle cluster in the null space of this operator will attach at least 2 zeros.
However this property is not true for the four particle states $T_6$, $T_4$, $T_3$,
$T_5$ or for the three particle $\LAngMom=0$ state because any state containing these terms will attach no zeros when three particles are brought close together.
Thus projecting out the three particle $\LAngMom=0$ state projects out these additional channels also.
The independent constraints can therefore be implemented by projecting out the following states
\begin{enumerate}
\item  3-particle channels of relative angular momenta $\LAngMom=0$.
\item  4-particle channels in the state $T_{22}$.
\end{enumerate}

The sequence of zeros attached to clusters of particles can be intuitively seen from the $\Lambda$-level structure of the composite fermions in the $2/3$ state. More specifically,
from the constraint that the composite fermions must compactly occupy the two lowest $\Lambda$-levels. 

Note that the $N$-boson state wherein all the bosons occupy the $m=0$ state,
corresponds to the state where the composite fermions occupy the lowest angular momentum states of the lowest $N$ $\Lambda$-levels,
\ie the states $(\Lambda=1,m=0),(\Lambda=2,m=-1)\dots (\Lambda=N,m=1-N)$.
This, for the case of $N=3,4$ and $5$ is shown on the left hand side of the three rows in the Fig.~\ref{fig:CFstructure}.
Also note that while the angular momenta of the composite fermions are not the same as those of the bosons, the changes made in angular momenta are the same.

Now consider the case where there are three bosons located close to the origin, in the $m=0$ state.
This is represented by the top-left panel of Fig.~\ref{fig:CFstructure}. In order to convert the state into the $2/3$ state,
the third particle, which is in the $\Lambda=3$ level, needs to be brought to the second $\Lambda$ level.
However, due to the fermion like exclusion by the two remaining particles that are already occupying $\Lambda=1,2$ orbitals,
the third particle needs to be moved to an orbital with $+2$ angular momentum,
equivalently the boson should be placed in the lowest Landau level orbital of angular momentum $2$, \ie in the orbital $\sim z^2_3$.
Thus the particle sees two zeros in the vicinity of the origin.
Note that these zeros arise as a result of exclusion by the two other particles and,
therefore, if the two particles are moved to another location,
the zeros will move with them rather than stay at the origin.
It is interesting to note that the complex correlations between the bosons, which effectively are complex many body constraints,
are reflected as simple fermi exclusion constraint for composite fermions. 
The cluster of the $3$ bosons has a total angular momentum of $2$,
which explains the absence of total relative angular momentum $0$ state of the three particles in the CF state.

Consider now the case of four particles shown in Fig. \ref{fig:CFstructure} second row. When the fourth boson is in the angular momentum zero state,
the corresponding composite fermion is in the $\Lambda=4$ level, and in the angular momentum $-3$ state.
In order to bring this composite fermion into the first two $\Lambda$-levels,
the fourth particle needs to be placed in the $\Lambda=1$ and $m=1$ orbital,
which requires an addition of $4$ units of angular momentum to the composite fermions,
and therefore also taking the boson from its $z^0$ (\ie\, $m=0$) state to the $z^4$ (\ie\, $m=4$) state.
Again by the arguments similar to the previous case,
this implies that the boson sees $4$ zeros in the cluster of $3$ other particles.
The total angular momentum of the cluster of these $4$ particles is $2+4$.
Given the $\Lambda$-level constraint and the fermi exclusion,
there is no way to reduce this angular momentum of the cluster.
Thus the minimum relative angular momentum of the cluster of four particles is $6$.
This explains the absence of four particle clusters with total relative angular momentum $\LAngMom\leq 5$ in the CF states.
These arguments also give an upper bound on the single particle angular momenta of the most compact clusters (smallest radius \ie smallest maximal single particle angular momentum).
The most compact cluster is obtained by placing the fourth boson in the angular momentum $4$ state.
This implies that the most compact cluster of four particles does not contain a single particle angular momneum $6$ state.
This explains the absence of the $T_6$  state in the $2/3$ wave function.

Similar arguments, summarized in the third row of Fig. \ref{fig:CFstructure} shows that the fifth particle sees $5$ zeros in the cluster of the remaining $4$ particles.
Further analysis along the same lines reproduces the vanishing sequence for the $2/3$ state $(0,0,2,4,5,7,8\dots)$.

For $\NSDTotal=4$ particles, we can explicitly perform the LLL projection and expand the bosonic/fermionic CF state wave function in terms of translationally invariant symmetric polynomials. The $2/3$ state of four particles is given by 
\begin{eqnarray}
	\Psi^{(B)}_{2/3}= \frac1{\sqrt{4945}}(4\sqrt{61} T_{42} - 63 T_{33}),
\end{eqnarray}
For the fermionic CF state, we have two different types of LLL projection,
Jain-Kamila projection (JKP) and direct projection (DP). Upto a Jastrow factor the $2/5$ state is given by 
\begin{eqnarray}
	\Psi^{(F)}_{2/5,JKP}&=& \frac1{\sqrt{3913}}(3\sqrt{61} T_{42} - 58 T_{33}),
	\\
	\Psi^{(F)}_{2/5,DP}&=& \frac1{\sqrt{17630}}(7\sqrt{61} T_{42} - 121 T_{33}).
\end{eqnarray}

Although the coefficients of $T_{42}$ and $T_{33}$ are slightly different in the different cases,
$T_6$ is absent in all cases.
Such universal absence supports the fact that the missing polynomials are consequences of CF $\Lambda$-level structures.

\section{Numerical construction of $T_\mu$ on the sphere}\label{sec:Num_costruction}
\label{sec:TmuOnSphere}
In order to numerically study the system, without being affected by edge effects, we will use the spherical geometry.
The bosons are confined to the surface of a sphere with a uniform radial magnetic field originating from a monopole of charge $2Q$
(in units of magnetic flux quanta) placed at the center.
In this section we will describe the states in the spherical geometry which are analogous to the translationally invariant polynomials $T_\mu$ in the disc geometry which were discussed in the previous sections. 

Translationally invariant polynomials $T_\mu$ in the disc geometry has the property that the polynomials are invariant under a constant translation $z_i=z_i+a$ for every particle $i$.
The center of mass angular momentum of any such state is $0$. Corresponding to each translation invariant state,
one can construct a sequence of center of mass excitations of these states by multiplying by $\left(\sum_{i=1}^\Nb z_i\right)^M$ where $M$ is the center of mass angular momentum. For any translation and rotational invariant Hamiltonian,
the energy eigenvalues are independent of the center of mass angular momenta.
Thus these sequence of states form an a degenerate multiplet.
Note that the center of mass excitations are not translationally invariant.

The combination of translational and rotational invariance of the Hamiltonian in the disc geometry corresponds to the full three dimensional rotational invariance on the sphere.
As a result, eigenstates are orbital angular momentum eigenstates and states that differ in the $L_z$ quantum number form a degenerate multiplet.
Angular momentum raising/lowering ($L_+/L_-$) in the sphere,
therefore corresponds to increase/decrease in center of mass,
in the corresponding scenario in the plane.
Translationally invariant states have the lowest center of mass angular momenta and therefore correspond to the highest weight states in the spherical geometry.
The full angular momentum multiplet can be obtained by sequentially using the raising operator $L_+$ on this state.

This argument can be applied to the special case of a single particle to show that an angular momentum $m$ single particle state in the disc,
correspond to an angular momentum $m-Q$ state in the sphere.
This can be seen from the fact that a translationally invariant state in the disc is a zero angular momentum state and center of mass excitations are the states $z^m$.
In the case of the spherical geometry these correpond to the highest weight state with angular momentum $m=-Q$ and the remaining $2Q$ states in its angular momentum multiplet. 

The correspondance between the single particle states,
implies that an $\Nsphere$ particle translationally invariant state of total relative angular momentum $\LAngMom$ on the disc 
corresponds to a highest weight state of total angular momentum $|\LAngMom-\Nsphere Q|$ on the sphere.

For instance, the Hamiltonian that projects out the relative angular momentum $\LAngMom=0$ states for any three particles can be written as
\begin{equation}
H = \sum_{i<j<k=1}^\Nsphere P_{ijk}(\LLockZero=3Q)
\end{equation}
$P_{ij\dots}(\LLockZero)$ is the projector on to angular momentum $\LLockZero$ multiplet of the particles $i,j\dots$, which is given by 
\begin{equation}
P_{ij\dots}(\LLockZero) = \sum_{m=-\LLockZero}^{\LLockZero} \left | \psi^{ij\dots}_{(\LLockZero,m)}\right \rangle \left \langle \psi^{ij\dots}_{(\LLockZero,m)} \right |
\end{equation}
where $\psi^{ij\dots}_{(\LLockZero,m)}$ is the state of the particles $ij\dots$ of total angular momentum $\LLockZero$ and a $z$-component $L_z=m$.
In case there is more than one multiplet of the same $\LLockZero$, the projector requires additional labels. For instance,
in the case of the $\LLockZero=4Q-4$ (corresponding to $\LAngMom=4$ of the disc),
there are two orthogonal states which correspond to the $T_{4}$ and $T_{22}$ in the disc. In the case of $\LLockZero=4Q-6$,
there are three states, corresponding to $T_{6}$, $T_{42}$ and $T_{33}$ in the disc.

The highest weight states corresponding to $T_{\mu}$ can in principle be obtained by constructing the space parallel and perpendicular to $P_{\mu}$, as was done in the case of the disc geometry.
Equivalently they can be constructed using the zero sequences (Table \ref{tab:zeros_of_T}). For instance,
$T_6$ does not vanish if three particles are brought close together,
whereas $T_{42}$ and $T_{33}$ do vanish.
Therefore $T_6$ is the only non zero eigenvalue state of the operator $\sum P_{ijk}(\LLockZero=3Q)$ in the $\LLockZero=4Q-6$ angular momentum space. The states $T_{42}$ and $T_{33}$ span the null space.
$T_{33}$ can be separated from $T_{42}$ by using the fact that $T_{33}$ is in the null space of the projector into the three particle states of angular momenta $\LLockZero=3Q-2$.
$T_{42}$ is then the vector orthogonal to $T_{33}$ and $T_6$. 

The states $T_{33},T_{42},T_{6}$ can also be identified by projecting out single particle states.
$T_{33}$ is the only state in the null space of the projector onto all single particles states of $\LLockZero=Q-6$ and $\LLockZero=Q-4$. In the two dimensional space orthogonal to $T_{33}$,
$T_{42}$ is the only null vector of the projector on to $\LLockZero=Q-6$.
It was checked numerically that the two approaches - (1) using the zero sequence and (2) single particle angular momenta - give the same results for $T_{\mu}$.

\begin{figure}
\vspace{2mm}
\includegraphics[width=\columnwidth]{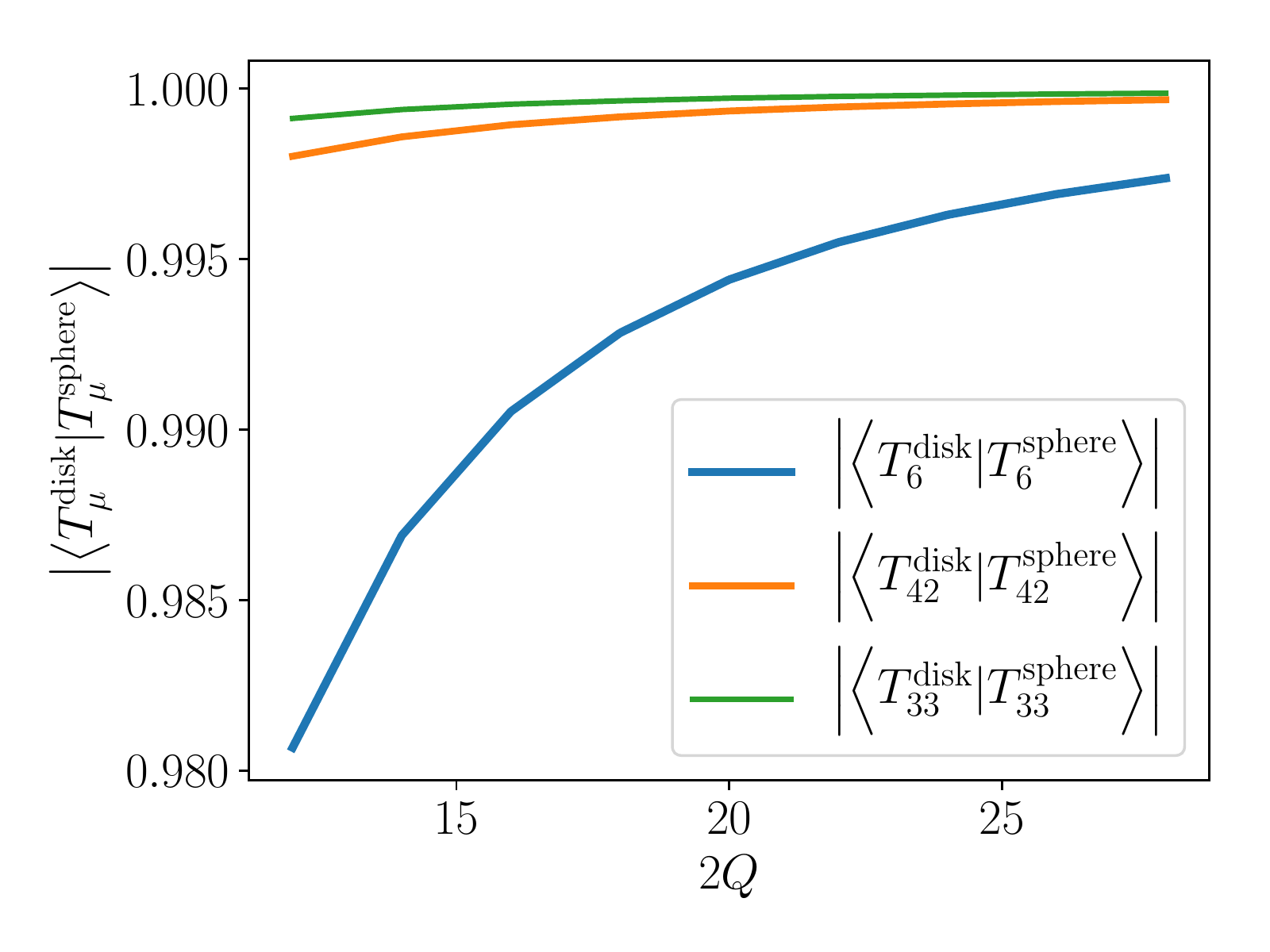}
\caption{Comparison of coefficients of $T_\mu$ (when expanded in monomial symmetric polynomials) in the disc and the spherical geometry. At finite  $Q$,
$\mu=6$ ($\mu=33$) shows the maximum (minimum) deviation from the disc geometry coefficients.
\label{fig-compare-disc-sphere}
}
\end{figure}

In the limit of large $Q$,
it is expected that the heighest weight states in the $T_{\mu}$ multiplets approach the ones obtained via stereographic projection from the disc.
To check that this indeed is the case,
we compared the coefficients in the expansion of $T_\mu$ in terms of the monomial symmetric polynomials in the disc and spherical geometry.
We considered the state $T^{\rm disk}_\mu$ in the disc geometry and the corresponding state in the spherical geometry $T^{\rm sphere}_{\mu}$ given by 
\begin{eqnarray}
T^{\rm disk}_\mu = \sum_{\lambda} c^{\rm disc}_\lambda P_\lambda\nonumber\\
T^{\rm sphere}_\mu = \sum_{\lambda^\prime} c^{\rm sphere}_{\lambda^\prime} P_{\lambda^\prime}
\end{eqnarray}
where $\lambda^\prime_i = \lambda_i-Q$ and calculated 
\begin{equation}
\left \langle T^{\rm disk}_{\mu}|T^{\rm sphere}_{\mu}\right \rangle \stackrel{\rm def}{=}\sum_\lambda {c^{\rm disc}_\lambda c^{\rm sphere}_{\lambda^\prime}}.
\end{equation}
The coefficients $c_{\lambda}$'s are defined such that $T_\mu$ and $P_\lambda$ are normalized.
As shown in Fig~\ref{fig-compare-disc-sphere}, this quantity indeed approaches $1$ for each $\mu$ in the limit of $Q\to \infty$. 
The difference between the coefficients in the sphere for finite $Q$ and the disc is a consequence of the curvature.
With increasing $Q$, the effect of curvature vanishes and the coefficients approach each other.
The state $T_{33}$ is the most compact and $T_{6}$ is the least.
As a result the latter shows the maximum deviation from disc coefficients due to curvature,
as can be seen in Fig~\ref{fig-compare-disc-sphere}.

\section{$\NSDGroup$-particle constraints}\label{sec:N_part_constraint}

\begin{table}

\begin{tabular}{|c|c|c|c|c|c|c|}
\hline state $\downarrow$
 & \multicolumn{3}{c|}{$3Q-\LLockZero=0$} & \multicolumn{3}{c|}{$3Q-\LLockZero=2$}\tabularnewline
\hline 
 $\Nb\rightarrow $ & $6$ & 8 & 10 & 6 & 8 & 10\tabularnewline
\hline 
\hline 
Exact & 0 & 0 & 0 & 0.40 & 0.87 & 2.23 \tabularnewline
\hline 
JK & 0 & 0 & 0 & 0.29 & 0.67 & 1.73\tabularnewline
\hline 
Coulomb & $4.1\times10^{-5}$ & $2.7\times10^{-4}$ & $9.51\times 10^{-4}$ & 0.63 & 1.18 & 3.08\tabularnewline
\hline 
\end{tabular}
\caption{Numerically obtained expectation values of the three particle projector $\sum_{i<j<k} P_{ijk}(\LLockZero)$ for $\LLockZero=3Q-0$ and $3Q-2$ for the three states: (i) CF state obtained by performing exact orthogonal projection into the lowest Landau level,
(ii) CF state obtained by Jain Kamila projection and (iii) ground state of the lowest Landau level Coulomb interaction.
Different columns indicate the values for $\Nsphere=6,8$ and $10$. 
\label{tab:3ParticleProjExpectation}}
\end{table}
In this section we numerically show that the $N$-particle constraints derived earlier are indeed satisfied by the CF state as well as the Coulomb ground state at filling fraction $2/3$. Expectation value of the projector
\begin{equation}
\mathcal{P}_{\NSDGroup}(\LLockZero) = \sum_{i_1,i_2\dots i_\NSDGroup}P_{i_1,i_2\dots i_\NSDGroup}(\LLockZero)
\end{equation}
can be used to check if an angular momentum $\LLockZero$ of $\NSDGroup$-particle sector is forbidden in a given state.
Here $P_{i,j,\dots}(\LLockZero)$ is the projector into the angular momentum $\LLockZero$ sector of the particles ${i,j,\dots}$. For instance,
the absence of three particle states of relative angular momentum $\LAngMom=0$ in the CF state implies vanishing of the expectation value of $\mathcal{P}_3(\LLockZero=3Q-0)$.
These expectation values are shown in the table \ref{tab:3ParticleProjExpectation} for different states as well as different finite size systems of $\Nb$ bosons.
The constraint is satisfied both by the CF state obtained by exact projection and Jain-Kamila projection. 
To a good approximation, the same constraints are satisfied also by the Coulomb ground state.
In contrast, the $\LAngMom=2$ sector of three particles is allowed and this results in a finite expectation value of $\mathcal{P}_3(\LLockZero=3Q-2)$.

\begin{table}

\begin{tabular}{|c|c|c|c|c|c|c|c|c|c|}
\hline 
\multicolumn1{|c}{$4Q-\LLockZero\to$} & \multicolumn1{c|}{} & 0 & 2 & 3 & 4 & 5 & \multicolumn{3}{c|}{$6$}\tabularnewline
\cline{8-10} 
\multicolumn1{|c}{} & $\Nb\downarrow$ &  &  &  &  &  & $T_{6}$ & $T_{42}$ & $T_{33}$\tabularnewline
\hline 
\hline 
\multirow{3}{*}{Exact} & $6$ & 0 & 0 & 0 & 0 & 0 & 0 & $0.27$ & $2.7$\tabularnewline
\cline{2-10} 
 & $8$ & 0 & 0 & 0 & 0 & 0 & 0 & $0.51$ & 4.7\tabularnewline
\cline{2-10} 
 & $10$ & 0 & 0 & 0 & 0 & 0 & 0 & 1.2 & 10.2\tabularnewline
\hline 
\multirow{3}{*}{JK} & $6$ & 0 & 0 & 0 & 0 & 0 & 0 & $0.19$ & $2.8$\tabularnewline
\cline{2-10} 
 & $8$ & 0 & 0 & 0 & 0 & 0 & 0 & $0.39$ & 4.8\tabularnewline
\cline{2-10} 
 & $10$ & 0 & 0 & 0 & 0 & 0 & 0 & 0.91 & 10.4 \tabularnewline
\hline 
\multirow{3}{*}{Coulomb} & $6$ & 0 & 0 & 0 & 0 & 0 & $10^{-4}$ & $0.42$ & 2.5\tabularnewline
\cline{2-10} 
 & $8$ & $10^{-12}$ & $10^{-7}$ & $10^{-5}$ & $10^{-3}$ & $10^{-4}$ & $10^{-4}$ & $0.69$ & 4.6\tabularnewline
\cline{2-10} 
 & $10$ & $10^{-6}$ & $10^{-6}$ & $10^{-6}$ & $10^{-3}$ & $10^{-7}$ & $10^{-4}$ & 1.7 & 10.1\tabularnewline
\hline 
\end{tabular}
\caption{Numerically obtained expectation values of the projectors into four particles sectors of angular momemta $4Q-\LLockZero=0,2,3,4$ and $5$. 
As expected from the constraints derived from the vanishing properties of the wave function,
these are exactly zero for the $2/3$ CF state, independent of whether the exact or the Jain-Kamila projection is used.
The Coulomb state also shows small values of this expectation value.
The last three columns show the expectation value for the projectors into the three linearly independent states in the space of angular momentum $4Q-\LLockZero=6$.
\label{tab:4ParticleProjExpectation}}
\end{table}

Table \ref{tab:4ParticleProjExpectation} shows the results for calculations in the $\NSDGroup=4$ particle sector.
In agreement with the constraints derived from the root-partition for the CF state,
we find that there are no $4$ particle clusters in the angular momentum sectors $\LAngMom=4Q-\LLockZero=0,2,3,4$ and $5$ in the CF state as well as the Coulomb ground state to a good approximation.
Four particle clusters can be in the $\LAngMom=4Q-\LLockZero=6$ sector only as a linear combination of $T_{42}$ and $T_{33}$. 

\begin{figure}
\includegraphics[width=\columnwidth]{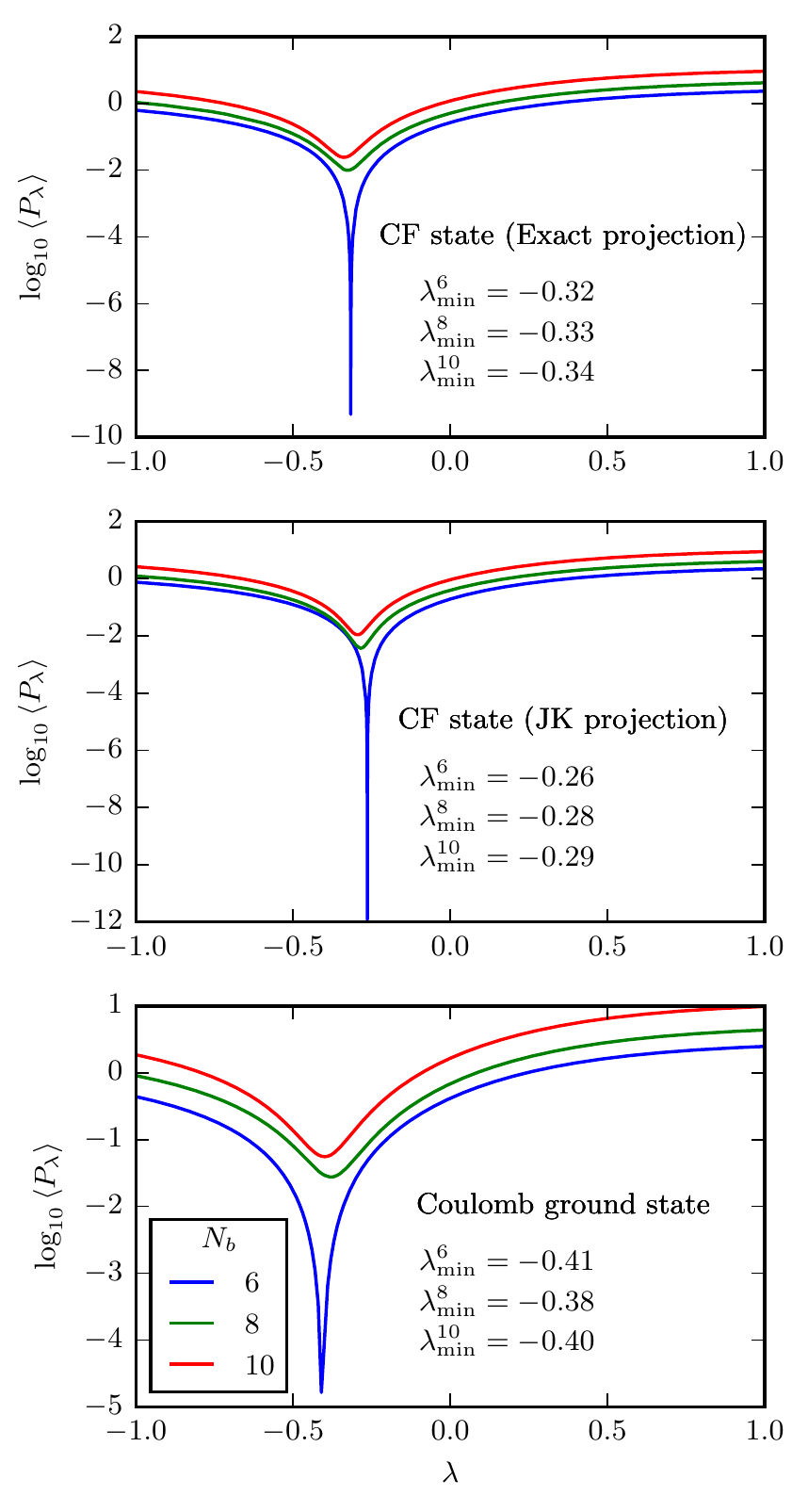}
\caption{
The expectation value of the projector $P_\lambda$ into the four particle sectors in a linear combination $T_{42}+\lambda T_{33}$. For $\Nb>6$,
there is no specific linear combintation that is fully projected out.
$\lambda_{\rm min}$ indicates the minima of the expectation value.
\label{fig:lambdaproj_L6}}
\end{figure}

The root partition of the CF state does not directly impose any constraint on the state of $4$ particle clusters in the $\LAngMom=6$ subspace.
To further understand the state of four particle clusters in the angular momentum $\LAngMom=4Q-\LLockZero=6$ channel,
we numerically find the expectation value of the operator defined as follows
\begin{equation}
P_{\lambda} = \sum_{i<j<k<l} P_{ijkl} (T_{42}+\lambda T_{33})
\end{equation}
where $P_{ijkl} (T_{42}+\lambda T_{33})$ is the projector into the state $T_{42}+\lambda T_{33}$ (parametrized by the mixing $\lambda$) of the four particles $i,j,k,l$.
If there is a specific linear combination $\lambda_0$ that is not allowed in the CF state,
the corresponding expectation value $P_{\lambda_0}$ should vanish.
Note that this is equivalent to all the four particle clusters in angular momentum $\LAngMom=6$ being in the orthogonal state $\lambda_0 T_{42}-T_{33}$. 

\begin{figure}
\includegraphics[width=\columnwidth]{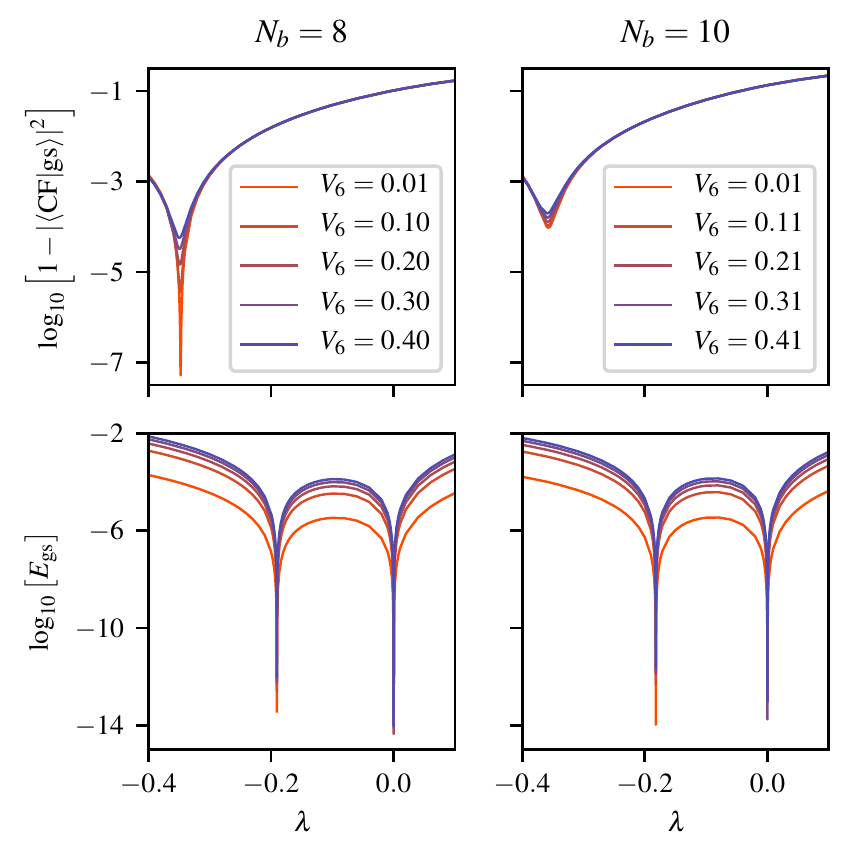}
\caption{
(top) Deviation from $1$ of the overlap-square of the CF wave function (obtained with exact projection) with the ground state of the constraint Hamiltonian (Eq. \ref{eq:constraint-Hamiltonian}),
  as a function of the mixing parameter $\lambda$.
  (bottom) The ground state energy of the Hamiltonian.
Different lines  show the coefficient $V_6$ of the projector onto $T_{42}+\lambda T_{33}$.
The left and the right panels show results for systems with $N=8$ and $10$ bosons, respectively.
\label{fig:gsprops}}
\end{figure}

Note that $\lambda$ is a constant and cannot be a function of any remaining coordinates,
as the angular momentum, \ie the order of the polynomial in $z_i$'s,
is $6$ and $T_{33}$ itself is of order $6$.
Figure \ref{fig:lambdaproj_L6} shows the numerically computed values of the expectation values $\left\langle P_{\lambda} \right \rangle$ (for the CF state and the Coulomb ground state) as a function of $-1<\lambda<1$.
We find that for $\Nb>6$ there is no $\lambda$ at which $\left\langle P_{\lambda} \right \rangle$ vanishes in the CF state. 

Figure \ref{fig:gsprops} shows the result of the diagonalization of the following Hamiltonian which,
in addition to projecting out $(\NSDGroup,\LAngMom)=(3,0)$ and $(4,4)$ projects out specific linear combinations $T_{42}+\lambda T_{33}$. For consistency of definitions in different systems, we will define $T_{33}$ and $T_{42}$ such that coefficients of $P_{33}$ and $P_{42}$ in $T_{33}$ and $T_{42}$ have opposite signs.
\begin{gather}
H_0 = \sum_{i<j<k} P_{ijk}(\LLockZero=3Q) + \sum_{i<j<k<l} P_{ijkl}(\LLockZero=4Q-4)\nonumber\\
H_1 = \sum_{i<j<k<l} P_{ijkl} (T_{42}+\lambda T_{33})\nonumber\\
H = H_0 +  V_6 H_1\label{eq:constraint-Hamiltonian}
\end{gather}
The spectrum of $H_0$ has multiple zero energy states, whose number increases with the number of particles.
A unique ground state is obtained upon adding $H_1$.
However, we find that for generic $\lambda$, the null spaces of $H_0$ and $H_1$ contain no common states.
As a result there is no zero energy state for generic $\lambda$.
The energy of the ground state depends on the strength  $V_6$ of $H_1$. For $\lambda=0$,
$H_1$ projects out the state $T_{42}$ state and Gaffnian is the zero energy ground state\cite{Simon07b}. 
Numerically, the root-partition of the state can be verified to be $003366\dots$.
We find an additional zero energy state at $\lambda \approx -0.19034$ for $\Nb=8$ ,
$\lambda\approx -0.18191$ for $\Nb=10$ and $\lambda\approx -0.17743$ for $\Nb=12$ whose root-partition is same as the root partition of the CF state at 2/3. 

It was reported in Ref.~\onlinecite{Regnault09} that there are $E[\frac{\Nb+2}{4}]$ $\LLockZero=0$ states in the space of states subdominant to the CF root-partition ($E[x]$ being the integer part of $x$).
For the system sizes considered here,
the numbers are $2,3$ and $4$ for $\Nb=8,10$ and $12$ respectively.
This implies that the imposing the constraints implied by the root-partition do not uniquely identify a state. However Eq.
\ref{eq:constraint-Hamiltonian} uniquely identifies a zero energy state with the CF root-partition.
This suggests that the constraint $H_1$ cannot be a consequence of constraints on the root partition.
However we find that this state is not identical to the CF state,
as can be inferred from the overlap of the state with the CF state (top panel of Fig \ref{fig:gsprops}). 

It is useful here to discuss briefly the numerical procedure used to determine the root partition.
Exact diagonalization gives the normalized ground states expanded in normalized states $P_\mu$ as $\psi=\sum_{\mu}c_\mu P_\mu$.
In general, due to the finite precision of our numerics, all coefficients will be nonzero. In addition,
error in determining the mixing parameter $\lambda$ leads to additional finite coefficients.
We searched for the dominant partitions among all the partitions $\mu$ with coefficients $|c_\mu| > 10^{-8}$ in the ground states.
The CF root partition $\alpha$ which has coefficient $|c_\alpha|$ of the order of $10^{-3},10^{-4},10^{-5}$ in the case of systems sizes $\Nb=8,10,12$,
dominates all partitions with coefficients greater than $10^{-8}$.
This can be made quantitative using the measure $1-\left|\left\langle {\rm gs} \left | P\right | {\rm gs} \right \rangle\right |$ where $P$ projects into all states that are subdominant to the CF root partition.
For the case of $N=8,10,12$ we find this number to be of the order of $10^{-14},10^{-15},10^{-15}$ respectively.

Fig \ref{fig:broaderlambda} shows the ground state energies for the Hamiltonian in a broader range of $\lambda$.
For $\Nb=10$, there appears to be two additional zero energy states.
However a clear root partition could not be obtained for the remaining states.

\begin{figure}
\includegraphics[width=\columnwidth]{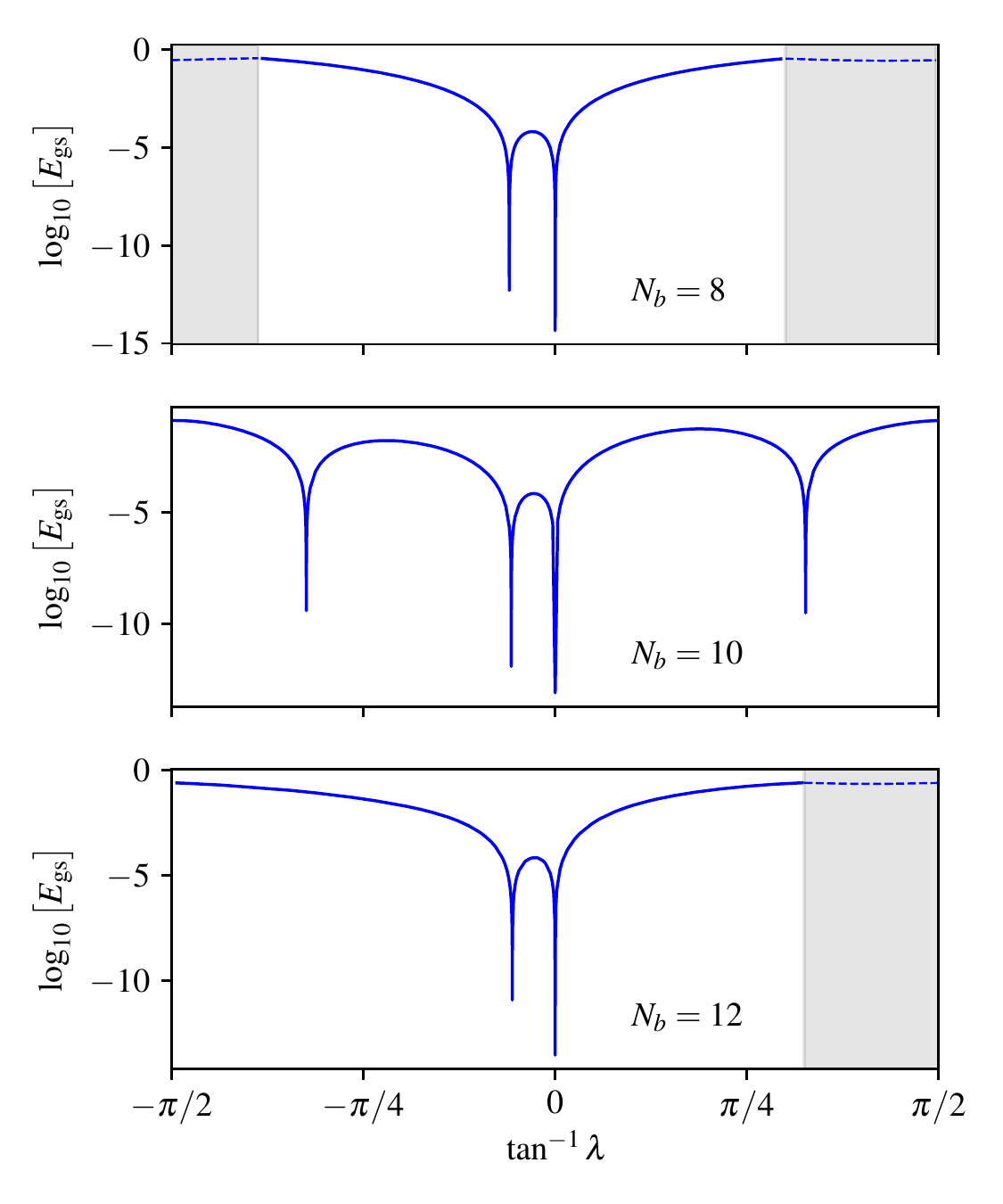}
\caption{Ground state energies of $\Nb=8,10$ and $12$ particles 
for the Hamiltonian in Eq \ref{eq:constraint-Hamiltonian} in broader regime $\lambda \in (-\infty,\infty)$ and for $V_6=0.2$.
The shaded region shows the parameters where the ground state has nonzero angular momentum.\label{fig:broaderlambda}}
\end{figure}

\vspace{1cm}

\section{Excitations spectra}\label{sec:excitations}
\begin{table}
\begin{tabular}{|l|l|l|}
\hline 
 & $(\Nb,2Q)$ & $L_{{\rm tot}}$ for $2/3$ CF states\tabularnewline
\hline 
\hline 
\multirow{3}{*}{gs (Fig \ref{fig:Neutral})} & (8,9) & 0(gs),2,3,4,5\tabularnewline
\cline{2-3} 
 & (10,12) & 0(gs),2,3,4,5,6\tabularnewline
\cline{2-3} 
 & (12,15) & 0(gs),2,3,4,5,6,7\tabularnewline
\hline 
\hline 
\multirow{3}{*}{3QP (Fig \ref{fig:QP} top)} & (9,9) & 1.5,2.5,4.5\tabularnewline
\cline{2-3} 
 & (11,12) & 0,2,3,4,6\tabularnewline
\cline{2-3} 
 & (13,15) & 1.5,2.5,3.5,4.5,5.5,7.5\tabularnewline
\hline 
\hline 
\multirow{3}{*}{2QP (Fig \ref{fig:QP} middle)} & (8,8 )& 0,2,4\tabularnewline
\cline{2-3} 
 & (10,11) & 1,3,5\tabularnewline
\cline{2-3} 
 & (12,14) & 0,2,4,6\tabularnewline
\hline 
\hline 
\multirow{3}{*}{1QP (Fig \ref{fig:QP} bottom)} & (9,10) & 3\tabularnewline
\cline{2-3} 
 & (11,13) & 3.5\tabularnewline
\cline{2-3} 
 & (13,16) & 4\tabularnewline
\hline 
\hline 
\multirow{3}{*}{1QH (Fig \ref{fig:QH} top)} & (7,8) & \multicolumn{1}{l|}{2}\tabularnewline
\cline{2-3} 
 & (9,11) & 2.5\tabularnewline
\cline{2-3} 
 & (11,14) & 3\tabularnewline
\hline 
\multirow{3}{*}{2QH (Fig \ref{fig:QH} bottom)} & (8,10) & \multicolumn{1}{l|}{0,2,4}\tabularnewline
\cline{2-3} 
 & (10,13) & 1,3,5\tabularnewline
\cline{2-3} 
 & (12,16) & 0,2,4,6\tabularnewline
\hline 

\end{tabular}
\caption{\color{black}Angular momentum quantum numbers $L_{\rm tot}$ for the CF $2/3$ state for the systems $(\Nb,2Q)$ 
for which the spectra shown in Figs \ref{fig:Neutral}, \ref{fig:QP} and \ref{fig:QH}. For the ground state (gs),
the $L_{\rm tot}=0$ corresponds to the ground state and the other values of $L_{\rm tot}$ to the neutral excitation (see text for details). The other results are for one,
two and three quasiparticles (1QP, 2QP and 3QP) and for one, two quasiholes (1QH, 2QH).
\label{tab:lowenergystates}}
\end{table}

\begin{figure*}
\includegraphics[width=\textwidth]{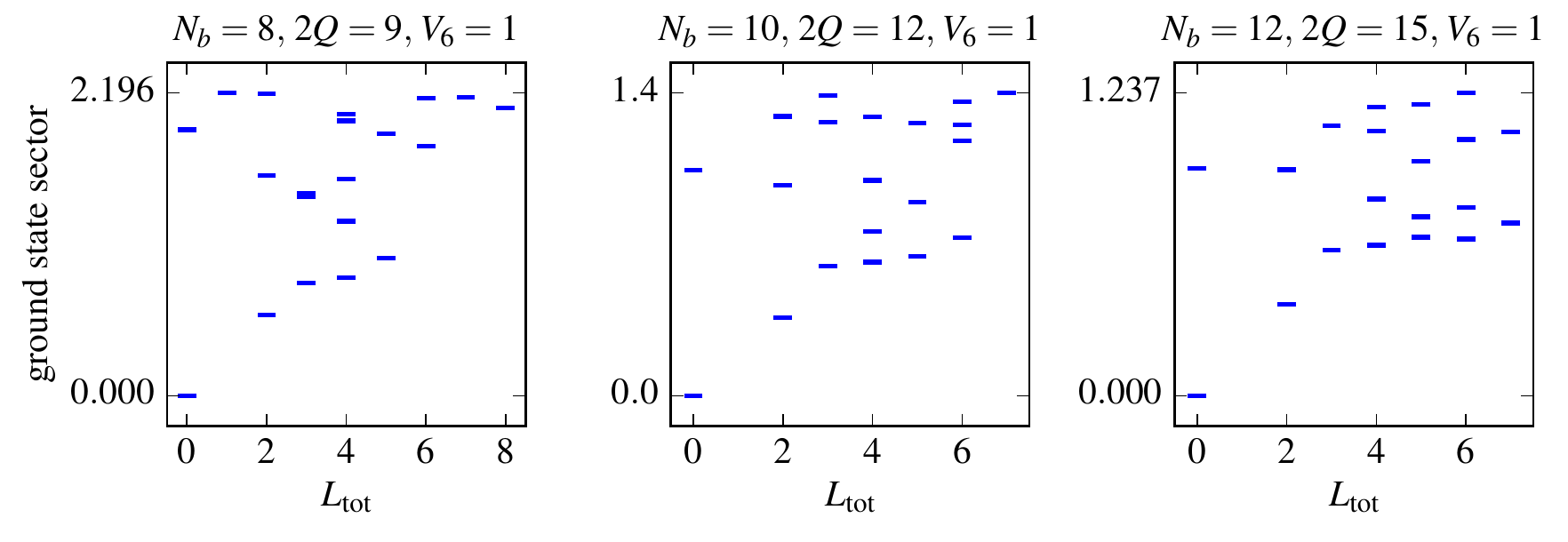}
\caption{Spectra of the Hamiltonian (Eq \ref{eq:constraint-Hamiltonian}) for systems in the ground state flux sector of the $2/3$ bosonic state. The panels show spectra for different system sizes.
$L_{\rm tot}$ indicates the total angular momentum quantum number of $\Nb$ bosons.
The counting of the neutral mode is same as in the 2/3 CF spectrum.
Note that for the case of the $2/3$ CF state the neutral mode states occur at $L_{\rm tot}=2,3,4\dots \frac{\Nb}{2}+1$.
The mixing parameter ($\lambda\approx 0.2$) in each case corresponds to the one where the zero energy state is found. 
\label{fig:Neutral}
}
\end{figure*}

To further explore the question of the relation between the $\lambda$-state and the CF state,
we consider excitations, which can be a very sensitive test of the nature of a state.
We consider the neutral excitations as well as quasiparticles and quasiholes,
obtained by subtraction and addition of flux to the system. In each case,
we choose the value of $\lambda$ corresponding to the zero energy ground state. 

The counting of low energy states as well as their quantum numbers is well understood for the CF states \cite{Dev92,Jain07,Balram13}. Within the CF theory,
the state of $\Nb$ bosons at flux $2Q$ maps into a state of $\Nb$ weakly interacting fermions at flux $2Q^*=2Q-N+1$.
The quantum numbers of the low-energy states can be obtained for the latter by simple angular momentum algebra. Consider, for example,
the neutral excitations. For the CF state for bosons at $\nu=2/3$,
we have $N_0=\Nb/2-1$ composite fermions in the lowest $\Lambda$ level and $N_1=\Nb/2+1$ composite fermions in the second $\Lambda$ level.
The lowest neutral excitation corresponds to the excitation of a composite fermion from the second to the third $\Lambda$ level,
creating a CF hole with angular momentum $l_h^*=\Nb/4$ and a CF particle with angular momentum $l_p^*=\Nb/4+1$.
This produces states at total angular momenta $L_{\rm tot}=2, 3, \cdots {\Nb\over 2}+1$,
where we have used that the state at $L=1$ is annihilated upon LLL projection\cite{Dev92}.
Table \ref{tab:lowenergystates} shows the quantum numbers of the low energy states for several systems as predicted by the CF theory.
We now give results for the model interaction that produces the $\lambda$-state as the exact zero energy ground state.

Figure \ref{fig:Neutral} shows the spectrum of the $\lambda$-state, obtained by diagonalizing the model Hamiltonian. 
The zero energy state is non-degenerate and has a finite gap separating it from a low energy mode  which has the same counting as the neutral modes of the $2/3$ CF state.

\begin{figure*}[h!]
\includegraphics[width=\textwidth]{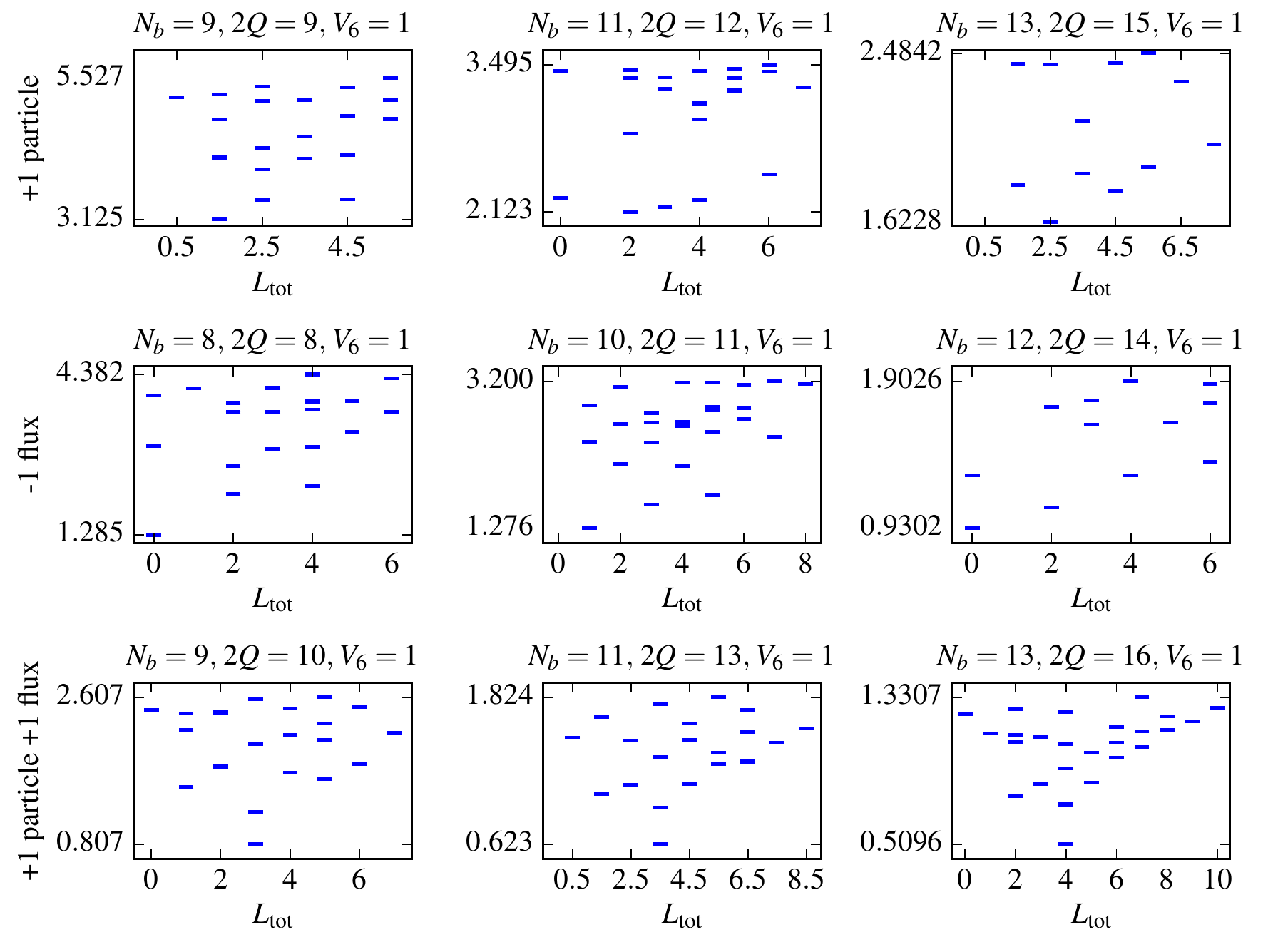}
\caption{\color{black}Spectra of the Hamiltonian in Eq~\ref{eq:constraint-Hamiltonian} for systems where we expect $2/3$ bosonic states with a few quasiparticles. Different columns indicate different system sizes.
In each case, the value of $\lambda$ is used at which a zero energy state was obtained in ground state sector of $\Nb=8,10,12$ particles, respectively. The top,
middle and bottom panels indicate systems where we expect three, two and one quasiparticles in the CF spectrum.
The angular momenta of the low lying states are same as those in the corresponding spectra of the $2/3$ CF state (Compare with Table \ref{tab:lowenergystates}).
\label{fig:QP}
}
\end{figure*}

Figure \ref{fig:QP} shows the low energy spectra for particle numbers and fluxes where we expect a few quasiparticles in the $2/3$ CF state. 
Addition of a particle to the system without changing the number of fluxes should result in three quasiparticles of the CF state.
The allowed angular momenta and the counting of low energy three-quasiparticle states matches exactly with the top panel of Fig \ref{fig:QP}. For instance for the $(\Nb, 2Q)=(13,15)$ system,
the three quasiparticles of the CF state should result in single low lying states at $L_{\rm tot}=1.5,2.5,\dots,5.5$ and $L_{\rm tot}=7.5$.
The low energy states are well separated from the bulk states for $V_6\sim 1$, but the gap decreases as $V_6$ decreases. 

Removal of a flux from the ground state sector without changing the number of bosons results in two quasiparticles of the CF state.
Again the counting of the two-QH state matches with the counting of the low energy states seen in the middle panels of Fig \ref{fig:QP}. 

Addition of a flux and a particle to the ground state flux sector gives a single quasiparticle of the CF state.
This should result in a single low energy state at angular momentum $L_{\rm tot}=\frac{1}{4}(\Nb+3)$.
This is indeed what we see in spectra  of the constraint Hamiltonian shown in the bottom panels of Fig \ref{fig:QP}.

\begin{figure*}
\includegraphics[width=\textwidth]{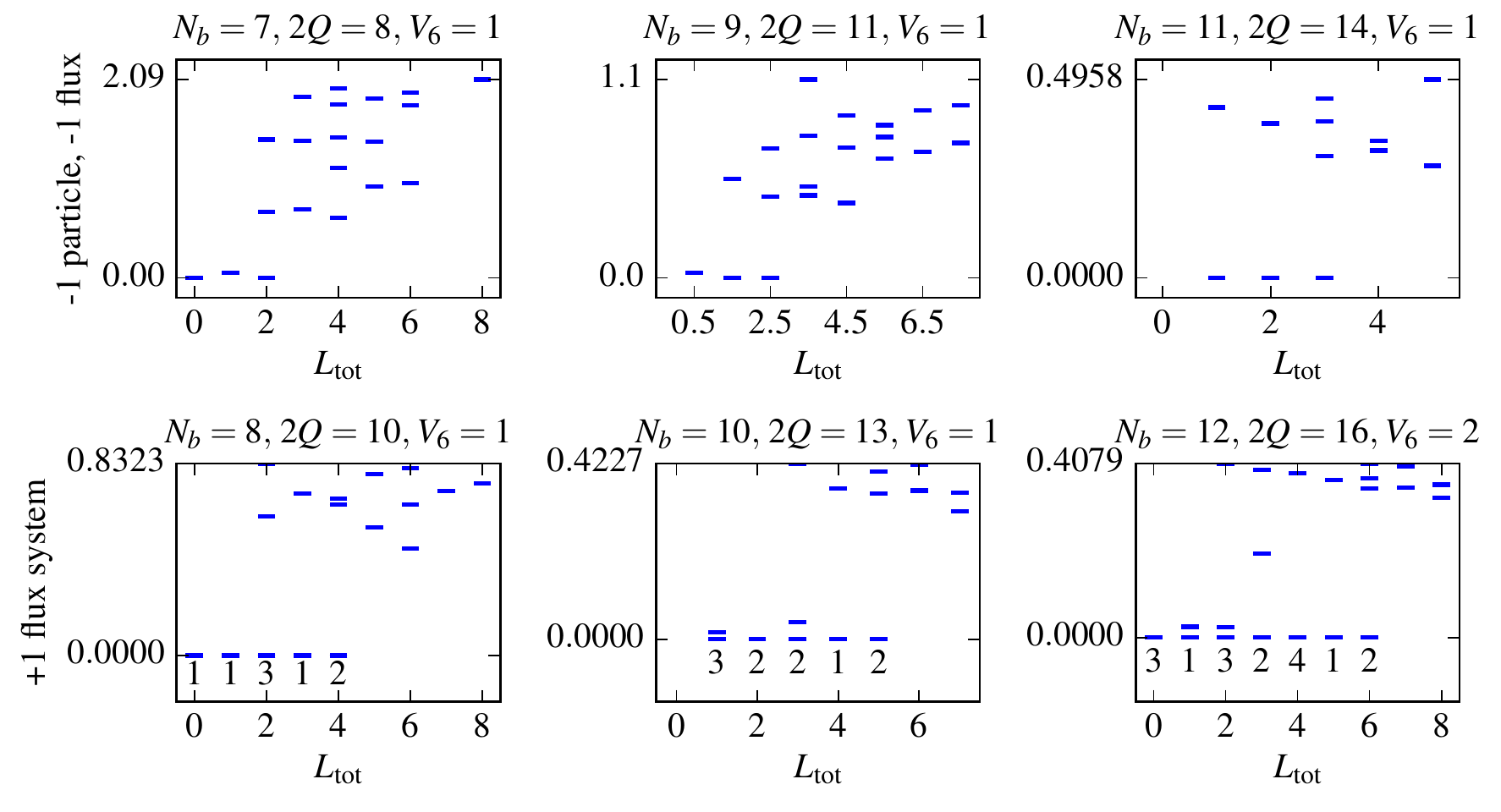}
\caption{\color{black}(top) Spectra of the Hamiltonian in Eq~\ref{eq:constraint-Hamiltonian} for systems where the CF theory would predict $2/3$ bosonic states with a quasihole at angular momentum quantum number $L_{\rm tot}=\frac{\Nb+1}{4}$. The spectra for our model interaction produces a very different structure with several nearly degenerate low-energy states. (bottom) Spectra for a system with one additional flux, which corresponds to two quasiholes for the $2/3$ CF state. Instead we see a proliferation of low energy states. The numbers next to the markers indicate the number of almost zero energy states (all depicted by the lowest dash) at that angular momentum.
\label{fig:QH}
}
\end{figure*}

\begin{figure*}
\includegraphics[width=\textwidth]{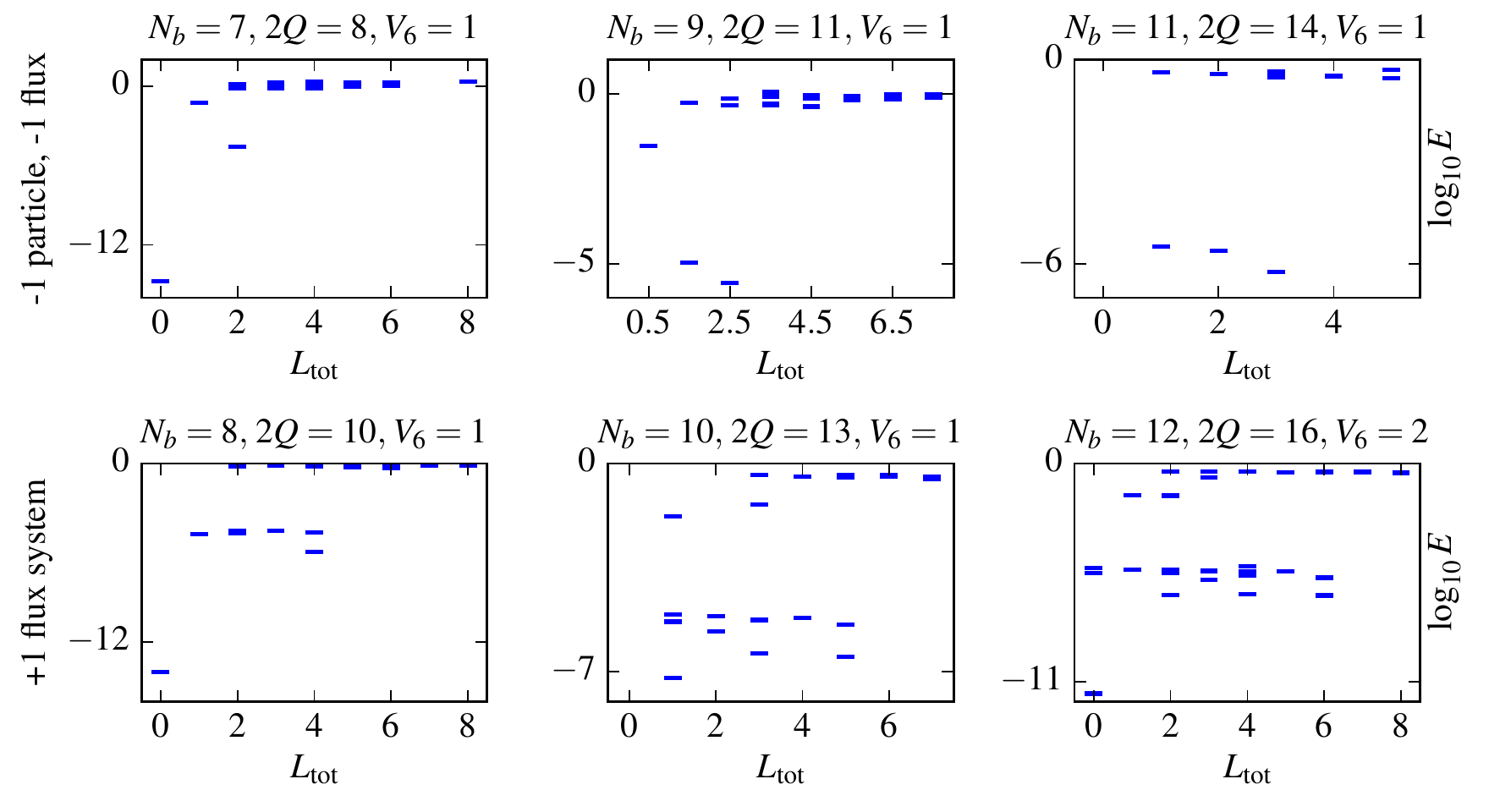}
\caption{\color{black}Same as in Fig \ref{fig:QH} but with the energy axis drawn in log-scale to reveal the structure of the states with almost zero energy. The lowest energy states for $\Nb>8$ have the same counting as expected from CF theory, but are not well separated from other states.\label{fig:QHlog}}
\end{figure*}

Removal of a particle and a flux should produce a single quasihole of the CF state. This should result in a single low energy state in the spectrum at angular momentum $L_{\rm tot}=(\Nb+1)/4$. The spectra of the constraint Hamiltonian for the corresponding system (Fig \ref{fig:QH}) show a qualitatively different structure. We expect a two-quasihole state upon addition of a flux to the ground state sector. The angular momenta of the corresponding low lying states in the CF spectra are given in Table \ref{tab:lowenergystates}. Spectrum of the constraint Hamiltonian instead shows a proliferation of almost zero energy states upon addition of flux. This could indicate that $\lambda$-state is potentially a gapless state in the thermodynamic limit.\cite{Read09} In both these cases, we find that the constraint Hamiltonian produces a significantly larger number of low energy states than the $2/3$ CF state. {Careful analysis of the spectra shows that for $\Nb >8$, the lowest energy states occur at quantum numbers predicted by the CF theory, but are not separated from other states by an identifiable gap. 
This can seen when the energy axis of the spectra are in a log-scale as shown in Fig-\ref{fig:QHlog}}

In summary, for quasiparticles as well as for neutral excitations, the constraint Hamiltonian produces the same counting of low energy states as the $2/3$ CF state.
For quasiholes, however, the constraint Hamiltonian produces more low energy states than the $2/3$ CF state. 

\section{Conclusions}

The initial motivation behind this work was to construct a local interaction for which the general CF wave functions are unique zero energy solutions. We considered the simplest non-trivial case, namely bosonic 2/3 FQHE for this purpose.
We have found that no interaction, including up to 4-particle interactions, obtains the CF 2/3 state as the exact and unique zero energy solution. 

We have, however, identified an interaction that produces a zero energy state, {called the $\lambda$-state}.
There are many indications that the $\lambda$-state is topologically equivalent,
to the CF state. The two have high overlaps, the same root partitions, and the same quantum numbers for the neutral and quasiparticle excitations. {However, their quasihole spectra do not produce a band of low energy states 
consistent with the QH band expected from the CF theory.} {The number of almost zero energy quasihole states rapidly increases with addition of fluxes, potentially indicating a gapless state in the thermodynamic limit.}

\begin{acknowledgments}
We thank H. Hansson, N. Regnault, A. Seidel, S. Simon and J.K. Slingerland for enlightening discussions.
We gratefully acknowledge financial support from Science Foundation Ireland Principal Investigator Award 12/IA/1697 (MF);
the National Research Foundation of Korea under Grant No. NRF-2018R1D1A1B07048749 (GSJ);
and the U. S. Department of Energy, Office of Basic Energy Sciences, under Grant no. DE-SC0005042 (JKJ).
\end{acknowledgments}

\pagebreak

\begin{appendix}

\section{Correspondence between disc and sphere}

We show two results in this appendix. First, wave functions that are annihilated by $L_{-}$ in the spherical geometry (i.e. are highest weight states with $L_z=-L$) produce, upon stereographic projection, disc wave functions that are translationally invariant. Second, the CF wave function of an integer number of filled $\Lambda$ levels in the sphere produces, after stereographic projection, the most compact disc wave function, i.e. the disc wave function that has the smallest size.

First let us define how to transform from sphere to disk. The stereographic projection corresponds to 
\begin{equation}
{u\over v}\rightarrow z
\end{equation}
The LLL single particle spherical wave function in general has the form
$$v^{2Q} f(u/v)=v^{2Q} f(z),$$ which implies that the many body wave function in the LLL has the form 
\begin{equation}
\prod_j v_j^{2Q} f(\{z_j\})
\end{equation}
We shall identify $f(\{z_j\})$ with the polynomial part of the disc wave function. In particular, the spherical Jastrow factor transforms into
\begin{equation}
\prod_{j<k}(u_iv_j-v_iu_j)^p = \prod_j v_j^{2pQ_1} \prod_j(z_j-z_k)^p
\label{Jastrow}
\end{equation}
where $2Q_1=N-1$ is the flux corresponding to filling factor 1.

Theorem 1: Wave functions that are annihilated by $L_{-}$ in the spherical geometry produce disc wave functions that are translationally invariant.

The proof is straightforward. Following the notation of Ref.~\onlinecite{Jain07}, let 
us now consider a spherical wave function that satisfies the condition
\begin{equation}
L_{-} \prod_j v_j^{2Q} f(z)=0   
\end{equation}
where 
\begin{equation}
L_{-}= -\sum_j  v_j  {\partial \over \partial u_j}
\end{equation}
is the angular momentum lowering operator. 
The condition $L_{-} \prod_j v_j^{2Q} f(z)=0$ is equivalent to 
\begin{equation}
\sum_k  {\partial \over \partial z_k}   f(z) =0
\end{equation}
which is precisely the condition for translational invariance 
$f(z_j+\eta)=f(z_j)$. QED.

To see how the FQHE wave functions transform, let us consider the $\nu^*=2$ state with two filled $\Lambda$ levels, which corresponds to 2/5 for fermions and 2/3 for bosons. In fact, in what follows we can consider any filling factor $\nu^*\leq 2$. 
Further, we will consider only states obtained from a single Slater determinant, so will drop multiplicative factors.

Let us recall the operators for the lowest $\Lambda$L:
\begin{equation}
Y_{Q^*,0,m}=u^{Q^*+m} v^{Q^*-m}
\end{equation}
and for the second $\Lambda$L:
\begin{equation}
Y_{Q^*,1,m}=u^{Q^*+m} v^{Q^*-m}[(Q^*+m+1)v \partial_v - (Q^*-m+1)u\partial_u]
\end{equation}
where $2Q^*$ is the effective flux for composite fermions.
A determinant formed from these operators on either a Jastrow factor (for bosons) or a Jastrow factor squared (fermions). We now make  transformation from $u,v$ to $z=u/v, v$, which corresponds to:
\begin{equation}
u\partial_u \rightarrow z \partial_z
\end{equation}
and 
\begin{equation}
v\partial_v \rightarrow -z\partial_z+v\partial_v
\end{equation}
In terms of the new variables we have
\begin{equation}
Y_{Q^*,0,m}=z^{Q^*+m} v^{2Q^*}
\end{equation}
\begin{equation}
Y_{Q^*,1,m}=z^{Q^*+m} v^{2Q^*}[(Q^*+m+1)v \partial_v - 2(Q^*+1)z\partial_z]
\end{equation}
Now, it turns out that the $v\partial_v \equiv 2pQ_1$ when applied to the Jastrow factor in Eq.~\ref{Jastrow}. 

The disc geometry wave function is thus a Slater determinant composed of 
\begin{eqnarray}
Y_{0,m}&=&z^{m}\nonumber\\
m&=&0, 1, \cdots 2Q^*
\end{eqnarray}
\begin{eqnarray}
Y_{1,m}&=&z^{m} [(m+1)2pQ_1 - 2(Q^*+1)z\partial_z]\nonumber\\
m&=&-1,\cdots 2Q^*+1
\end{eqnarray}
which acts on $\prod_{j<k}(z_j-z_k)^p$.

Let us consider the bosonic $\nu=2/3$, for which $Q^*=(N-4)/4$ and $p=1$ and we make a Slater determinant with {\em all} $Y_{0,m}$ and $Y_{1,m}$.  After eliminating many terms using row operations, we get the following form
\begin{widetext}
\begin{equation}
\chi_{2/3}=
\begin{vmatrix}
1 & 1 & \cdots \\
z_1 & z_2 & \cdots\\
.. & .. & \cdots\\
z_1^{2Q^*} & z_2^{2Q^*}  & \cdots \\
\partial_1 & \partial_2 & \cdots \\
z_1 \partial_1 & z_2\partial_2 & \cdots\\
.. & .. & \cdots\\
z_1^{2Q^*+1} \partial_1 & z_2^{2Q^*+1}\partial_2 & \cdots\\
2Q_1 z_1^{2Q^*+1}- z_1^{2Q^*+2} \partial_1 \;\;\;\;& 2Q_1 z_2^{2Q^*+1}- z_2^{2Q^*+2} \partial_2 \;\;\;\; & \cdots\\
\end{vmatrix} \prod_{j<k}(z_j-z_k)
\end{equation}
\end{widetext}
Note that the first term in the last row {\em cannot} be eliminated.

As proved before, this wave function is guaranteed to be translationally invariant.
What is the largest power of $z_1$ in this wave function? One may naively think that it is ${3\over 2}N-2$, which is the sum of $N-1$ from the Jastrow factor and $2Q^*+1={N\over 2}-1$ from the last row in the Slater determinant.  However, because 
\begin{equation}
[2Q_1 z_1^{2Q^*+1}-z_1^{2Q^*+2} \partial_1]z_1^{N-1}=0
\end{equation}
with $2Q_1=N-1$, this power is absent. In other words, the outermost occupied orbital has $m_{\rm max}={3\over 2}N-3$. If we write the wave function directly in the disc geometry with half of the composite fermions in the lowest $\Lambda$L and half in the second $\Lambda$L, we would get $m_{\rm max}=N-1+(N/2)-1={3\over 2}N-2$.

The same remains true for the fermionic 2/5 state with $p=2$
\begin{widetext}
\begin{equation}
\chi_{2/5}=
\begin{vmatrix}
1 & 1 & \cdots \\
z_1 & z_2 & \cdots\\
.. & .. & \cdots\\
z_1^{2Q^*} & z_2^{2Q^*}  & \cdots \\
\partial_1 & \partial_2 & \cdots \\
z_1 \partial_1 & z_2\partial_2 & \cdots\\
.. & .. & \cdots\\
z_1^{2Q^*+1} \partial_1 & z_2^{2Q^*+1}\partial_2 & \cdots\\
4Q_1 z_1^{2Q^*+1}- z_1^{2Q^*+2} \partial_1 \;\;\;\;& 4Q_1 z_2^{2Q^*+1}- z_2^{2Q^*+2} \partial_2 \;\;\;\; & \cdots\\
\end{vmatrix} \prod_{j<k}(z_j-z_k)^2
\end{equation}
\end{widetext}
Following in the same fashion as before, the outermost occupied orbital has $m_{\rm max}= {5\over 2}N-4$ (rather than the naive ${5\over 2}N-3$).

\end{appendix}

\bibliography{biblio_fqhe}
\bibliographystyle{apsrev}

\end{document}